\documentclass[a4paper, 12pt]{article}
\usepackage{a4}
\usepackage{amsfonts}
\usepackage{amssymb}
\usepackage{cite}
\usepackage{epsfig}
\usepackage{psfig}

\author{}
\title{Chiral D-brane Models with Frozen Open String Moduli}

\newcommand{\drawsquare}[2]{\hbox{%
\rule{#2pt}{#1pt}\hskip-#2pt%  left vertical
\rule{#1pt}{#2pt}\hskip-#1pt%  lower horizontal
\rule[#1pt]{#1pt}{#2pt}}\rule[#1pt]{#2pt}{#2pt}\hskip-#2pt%  upper horizontal
\rule{#2pt}{#1pt}}% right vertical
\newcommand{\fund}{\raisebox{-.5pt}{\drawsquare{6.5}{0.4}}}%  fund
\newcommand{\Ysymm}{\raisebox{-.5pt}{\drawsquare{6.5}{0.4}}\hskip-0.4pt%
         \raisebox{-.5pt}{\drawsquare{6.5}{0.4}}}%  symmetric second rank
\newcommand{\Yasymm}{\raisebox{-3.5pt}{\drawsquare{6.5}{0.4}}\hskip-6.9pt%
        \raisebox{3pt}{\drawsquare{6.5}{0.4}}}%  antisymmetric second rank
\newcommand{\antifund}{\overline{\fund}}

\newcommand{\be}{\begin{equation}}
\newcommand{\ee}{\end{equation}}
\newcommand{\bea}{\begin{eqnarray}}
\newcommand{\eea}{\end{eqnarray}}
\newcommand{\mbb}{\mathbb}
\newcommand{\ti}{\times}
\newcommand{\half}{\frac{1}{2}}
\def\IR{\relax{\rm I\kern-.18em R}}
\def\o#1{\overline{#1}}

\def\cn{{\cal N}}
\def\cm{{\cal M}}
\def\ca{{\cal A}}
\def\eps{\epsilon}
\def\T{{\bf T}}
\def\P{{\bf P}}
\def\K3{{\bf K3}}
\def\Om{\Omega}

\def\th{\theta}
\def\Th{\Theta}
\def\G{\Gamma}
\def\a{\alpha}
\def\b{\beta}
\def\d{\delta}
\def\i{\iota}
\def\sig{\sigma}
\def\car{{\cal R}}
\def\ov{\overline}
\def\preal{{\rm Re\,}}
\def\pim{{\rm Im\,}}
\def\raw{\rightarrow}
\def\longraw{\longrightarrow}

\long\def\symbolfootnote[#1]#2{\begingroup%
\def\thefootnote{\fnsymbol{footnote}}\footnote[#1]{#2}\endgroup}

\begin{document}

\title{
\begin{flushright} \vspace{-2cm}
{\small MAD-TH-05-1 \\
\small MPP-2004-115 \\
\small UPR-1091-T \\ \vspace{-0.35cm}
hep-th/0502095} \end{flushright}
\vspace{2.0cm}
Chiral D-brane Models with Frozen \\ Open String Moduli \\ \quad
}
\author{\small Ralph~Blumenhagen$,^\clubsuit$\, Mirjam~Cveti\v c$,^\heartsuit$\, Fernando~Marchesano$,^\spadesuit$\, and\, Gary~Shiu$^\spadesuit$}
\date{}

\maketitle

\begin{center}
\emph{$^\clubsuit$ Max-Planck-Institut f\"ur Physik, F\"ohringer Ring 6, 80805 M\"unchen, Germany } \\
\vspace{0.1cm}
\emph{$^\heartsuit$ Department of Physics and Astronomy, University of Pennsylvania,
 Philadelphia, PA 19104-6396, USA }\\
\vspace{0.1cm}
\emph{$^\spadesuit$ Department of Physics, University of Wisconsin,
  Madison WI 53706, USA\symbolfootnote[2]{Permanent address}} \\
\vspace{0.1cm}
\emph{and} \\
\vspace{0.1cm}
\emph{Perimeter Institute for Theoretical Physics, \\
Waterloo, Ontario N2L 2Y5, Canada}  
\\
\vspace{1.0cm}
\end{center}

\begin{abstract}
\noindent 
Most intersecting D-brane vacua in the literature contain additional massless adjoint fields in their low energy spectrum. The existence of these additional fields make it difficult to obtain negative beta functions and, eventually, asymptotic freedom. We address this important issue for $\cn=1$ intersecting D-brane models, rephrasing the problems in terms of (open string) moduli stabilization. In particular, we consider a $\mbb{Z}_2 \times \mbb{Z}_2$ orientifold construction where D6-branes wrap {\em rigid} 3-cycles and such extra adjoint fields do not arise. We derive the model building rules and consistency conditions for intersecting branes in this background, and provide $\cn=1$ chiral vacua free of adjoint fields. More precisely, we construct a Pati-Salam-like model whose $SU(4)$ gauge group is asymptotically free. We also comment on the application of these results for obtaining gaugino condensation in chiral D-brane models. Finally, we embed our constructions in the framework of flux compactification, and construct new classes of $\cn=1$ and $\cn=0$ chiral flux vacua.

\end{abstract}

\thispagestyle{empty}
\clearpage

\tableofcontents

\section{Introduction}

During the past few years, the long-standing problem of moduli stabilization in string theory has received much renewed attention. The resurgence of interest is due largely to the observation that compactifications with fluxes provide a systematic, calculable, and geometrical framework to stabilize closed string moduli \cite{gvw99,drs99,tv99,gks00,cklt00,gkp01,bb01,kst02}. Although not directly addressed in the original works cited above, it has recently been shown that these background fluxes also have  the additional effect of stabilizing open string moduli \cite{gktt04,cu04,ciu04}. From the effective field theory point of view, this stabilization of open string moduli can be understood as flux-induced soft terms in the gauge theories on the D-brane world-volumes. Thus, flux compactification has emerged as an attractive phenomenological scenario, since it provides a framework where moduli stabilization and supersymmetry (SUSY) breaking -- two central problems in string phenomenology -- can in principle be analyzed simultaneously in a controlled, stringy manner.

However, in order to explore {\it quantitative} features of this scenario, it is important to construct some concrete realistic models where explicit computations (and hopefully predictions) can be made. In particular, important issues, such as the pattern of soft SUSY breaking terms \cite{MG00,ciu03,ggjl03,lrs04,ciu04} and  the signatures and stability of cosmic strings (see \cite{GS02,TK04,JP04} for some reviews) formed at the end of brane inflation \cite{dt98,kklmmt03}, depend critically on how the Standard Model (SM) is embedded. In the past few years, the intersecting brane (and, in the T-dual picture, magnetized D-brane) world scenario \cite{bgkl00a,afiru00,afiru00a,bkl00,imr01,bklo01}  has proven to be a promising framework to construct semi-realistic, $D=4$, ${\cal N}=1$ chiral string models \cite{csu01,csu01a,bgo02,GH03,lp03,ho04}.\footnote{For reviews on this rich subject and a more complete list of references see \cite{AU03,DL04,RB04,bcls05}.} More importantly, it has recently  been realized \cite{ciu04,ms04,ms04a} that by turning on background NSNS and RR fluxes in magnetized D-branes models, SUSY can be {\it softly}  broken while satisfying the supergravity equation of motion (hence the usual problem of the classical instability in non-supersymmetric models due to non-vanishing NSNS tadpoles is absent). Elaborating on the framework developed in \cite{btl03,cu03}, some three-generation MSSM-like flux vacua have been constructed in \cite{ms04,ms04a}, thus providing the first concrete examples with these desirable features in string theory. These models also demonstrated, as a proof of concept, that string models with realistic particle physics features can be embedded in the framework of flux compactification.\footnote{Subsequently, flux vacua with semi-realistic features have been constructed in \cite{cl04,cll05}.}

When constructing semi-realistic D-branes models, one often focuses on obtaining a realistic gauge and chiral sector in the low energy spectrum of the theory, i.e., a spectrum free of SM chiral exotics. While this basic feature can indeed be achieved, any of the models constructed so far contain also some additional {\it non-chiral} open string sector states. To be concrete, let us consider $\cn=1$ chiral models based on intersecting D6-branes. In addition to the chiral spectrum localized at the D-brane intersections, these models typically possess non-chiral open-string states, usually associated with the D-brane positions and Wilson lines. If these open-string sector moduli are not completely frozen (which is the case of almost all the vacua built so far), one gets additional  adjoint or (anti)symmetric matter in the effective four dimensional theory, typically charged under the SM gauge group. Geometrically, the appearance of such moduli is a consequence  of the D-branes wrapping 3-cycles which are not completely rigid, and which then allow  for (D- and F-) flat deformations in the effective  (supersymmetric) theory.

Clearly, the  Standard Model of particle physics as well as its supersymmetric extension, the Minimal Supersymmetric Standard Model (MSSM), do not contain these light  scalars. In fact, one of the strongest motivations behind the MSSM is the apparent unification of gauge couplings at an energy scale $M_{GUT} \simeq 10^{16}$ GeV, a result which heavily relies on the light particle content of the MSSM. Any extra light  matter gives  an additional positive  contribution to the gauge coupling beta-function, and hence could easily spoil the celebrated properties of asymptotic freedom (for the strong interactions) and perturbative gauge coupling unification enjoyed by the MSSM. 

Besides the Standard Model sector, these D-brane constructions typically come equipped with a ``hidden sector'' in the low energy theory, where the mechanism of gaugino condensation can be invoked in order to provide an extra source of supersymmetry breaking. However, this mechanism also requires the corresponding gauge theory to be asymptotically free, thus also making it undesirable to have additional non-chiral hidden sector matter. Gaugino condensation has also recently been employed in the KKLT scenario \cite{kklt03} to obtain de Sitter vacua from  Type IIB string flux compactifications. Here a gaugino condensate on some D7-brane gauge theory provides a non-perturbative contribution to the $D=4$ effective superpotential, breaking the no-scale property of the flux induced superpotential and freezing the K\"ahler moduli. Again, the existence of additional non-chiral matter in the hidden sector makes it difficult to realize this celebrated mechanism.

Of course, the common wisdom is that these additional non-chiral fields (both in the observable and hidden sectors) will generically pick up a mass once SUSY is broken. This is nothing but the usual statement that moduli are expected to be lifted upon SUSY breaking. In the context of flux compactification, however, the background fluxes provide an explicit way of computing both moduli lifting and SUSY breaking. Hence, instead of resorting to some unknown (often non-perturbative) SUSY breaking mechanism, we can calculate explicitly the flux-induced masses acquired by the moduli, at least for simple toroidal orbifold backgrounds and neglecting the back-reaction of the D7-branes. 
Towards this end, let us note that: 

\begin{itemize}

\item 
Not all moduli are lifted by the fluxes. The K\"ahler moduli in the closed string sector, the D3 moduli and the D7 Wilson lines in the open string sector are in principle not stabilized by the background fluxes.

\item
Even for the geometric open string moduli that are lifted, like the adjoint fields associated with the position of D7-branes, the flux-induced masses will generically be of the same scale as the MSSM soft masses since they arise from the same source, i.e., the 3-form fluxes.

\end{itemize}

In most SUSY scenarios, the soft masses are of the order of the TeV scale.\footnote{An exception is split SUSY \cite{ad04} where the soft masses are well above the TeV scale. However, a different hierarchy (between the soft masses and the gaugino masses) needs  to be generated to preserve gauge unification.} Therefore, even if these additional geometric open string moduli could acquire a mass from the fluxes, it would typically be of the TeV scale and hence still pose a problem for gauge unification. In principle, one could generate a hierarchy of scales between the MSSM soft masses and the masses of these additional matter by means of strongly warped throats. This might indeed work for the hidden sector open string moduli. For instance, one could think of the hidden sector branes and the Standard Model branes being located at positions in the extra dimensions where the warp factors are drastically different. However, it seems difficult for such a warp factor to generate a hierarchical difference between the MSSM soft masses and  these additional open string moduli charged under the Standard Model. In fact, the above argument is quite general, and is not limited to the specific mechanism of moduli stabilization/SUSY breaking. The existence of open string moduli charged under the SM gauge group poses a new hierarchy problem: one needs to generate a hierarchical difference between the masses of fields on the same stacks of D-branes.

A way out of this conundrum is to get rid of these geometric open string moduli from the very beginning. In this paper, we explore this possibility by constructing chiral  ${\cal N}=1$ D-brane models with the associated D-branes wrapped around {\it rigid} cycles. Since the cycles are rigid, there are simply no moduli associated with the positions of the D-branes or, more precisely, such would-be moduli have a string scale mass.

To illustrate this class of constructions, we will focus on type II string theory compactified on toroidal orientifold backgrounds. These class of models have indeed been the predominant source of explicit compact examples\footnote{Besides orbifold constructions, semi-realistic models based on non-geometric conformal field theory, such as orientifolds of  Gepner models \cite{abpss96,bw98}, have most recently  been considered in \cite{aaln03,RB03a,bhhw04,bw04,dhs04,aaj04,bw04a,dhs04a}. These vacua are known to describe special points in the moduli space of certain Calabi-Yau manifolds, and the results in the literature indicate that supersymmetric semi-realistic chiral particle spectra can be obtained in many of these models \cite{dhs04a}. Since the K\"ahler moduli at the Gepner points are of stringy size, it is difficult to analyze the effects of, e.g, non-trivial background fluxes.} involving intersecting/magnetized D-branes and, due to their simplicity, they allow to easily compute quantities of the low energy theory which do not have a topological description like, e.g., the masses of vector-like pairs and other non-chiral states. More precisely, we will consider an orientifold of type IIA string theory compactified on $\T^6/(\mbb{Z}_2 \times \mbb{Z}_2'$). As we will see, this background naturally involves rigid 3-cycles that intersecting D6-branes can wrap, and allows to construct chiral models where almost every open string modulus is absent. Our purpose here is to derive the consistency conditions and develop the string model building techniques that allow to construct chiral $\cn=1$ vacua in this framework, as well as provide some explicit semi-realistic examples that illustrate the ideas above. Moreover, we will briefly discuss some phenomenological aspects of such D-brane models.

This paper is organized as follows. In Section \ref{moduli} we describe the issue of open string moduli stabilization in $\cn=1$ intersecting D-brane constructions. In particular, we briefly overview the $\cn=1$ chiral models constructed so far, showing that they were made of D6-branes wrapping non-rigid 3-cycles.  The reader not interested in this general discussion may start directly from Section \ref{intersecting} where, motivated by these facts, we consider a toroidal orientifold which contains a large set of rigid 3-cycles. More precisely, we study type IIA theory on the $\T^6/(\mbb{Z}_2\times \mbb{Z}_2)$ orientifold with discrete torsion. Notice that (a T-dual version of) this orientifold background has been previously studied in \cite{aaads99}, as an example of the {\em brane supersymmetry breaking} orientifolds. One would naively think that no $\cn=1$ D-brane vacua can be constructed in this background. Following \cite{ms04a}, we show that this is not the case and, in Section \ref{examples}, construct explicit $\cn=1$ chiral models from intersecting rigid D6-branes. In particular, we construct a Pati-Salam type model with four generations of chiral matter. These constructions illustrate the general framework where more realistic models can in principle be obtained. In Section \ref{gaugino} we show how, by using D-branes wrapped on rigid cycles, asymptotically free gauge theories on the worldvolume of D-branes can be constructed. Finally, in Section \ref{flux}, we embed our constructions in the framework of flux compactification, and construct new classes of ${\cal N}=1$ and ${\cal N}=0$ chiral flux vacua. We end with some conclusions in Section \ref{conclusions}. Some details about K-theory constraints of these models are relegated to the Appendix.

After this paper was completed and prepared for submission, we noticed \cite{dt05} which, in a different spirit, also studies some chiral models on a similar $\mbb{Z}_2 \ti \mbb{Z}_2$ background.

\section{Open string moduli for intersecting branes}\label{moduli}

In this section we describe the issue of open string moduli in intersecting D-brane constructions from a general viewpoint, in order to motivate our search for chiral D-brane models in specific backgrounds. As usual, such a problem is easier to understand in supersymmetric constructions, on which we henceforth focus. In order to illustrate the problem of open string moduli stabilization, we briefly describe the global $\cn =1$ chiral models based on intersecting D6-branes constructed so far, which are living in  toroidal orientifold backgrounds. We point out why in general one would expect to have adjoint matter fields in this  class of constructions. We also describe our strategy for getting rid of the adjoint fields by means of D6-branes wrapped on {\em rigid} 3-cycles.

\subsection{Open string moduli on Calabi-Yau's}

Let us consider type IIA string theory compactified on a six-dimensional manifold $\cm_6$. An intersecting D6-brane model in this context is given by a collection of D6-branes filling up the four non-compact dimensions and wrapping some 3-cycles $\Pi_a \subset \cm_6$. If $\Pi_a$ contains $N_a$ coincident D6-branes, the gauge group that we obtain is $U(N_a)$, so we will usually have a gauge theory of the form $\prod_a U(N_a)$. The chiral matter of this theory can be obtained from the homology classes $[\Pi_a] \in H_3(\cm_6,\mbb{Z})$ and their intersection numbers $I_{ab} = [\Pi_a] \cdot [\Pi_b]$. Indeed, the latter encode the multiplicity and the chirality of the fermions transforming in the bifundamental representation $(\ov{N}_a, {N}_b)$.

Let us now restrict to supersymmetric backgrounds, i.e., those which admit a covariantly constant spinor. If we restrict to closed string backgrounds without field strength fluxes turned on, we recover the usual result that $\cm_6$ must be a Calabi-Yau three-fold. Such a manifold comes equipped with a holomorphic 3-form $\Om_3$, which indicates which are the supersymmetric 3-cycles of the theory. Indeed, $N_a$ D6-branes wrapping $\Pi_a$ will yield a supersymmetric $U(N_a)$ gauge theory if $\Pi_a$ is calibrated by the 3-form $\preal (e^{i\phi_a} \Om_3)$. Such a 3-cycle is named Special Lagrangian (SL) with phase $\phi_a$, and it minimizes its volume in the homology class $[\Pi_a]$ which, being a topologically invariant, encodes the RR charges of the D6-branes on $\Pi_a$.

In this particular setup it is easy to understand which are the open string moduli of the compactification. Indeed, McLean's theorem \cite{McLean} states that the moduli space of deformations of a compact SL $\Pi_a$ is a smooth manifold of real dimension  $b_1(\Pi_a)$. String theory complexifies this moduli space by adding $b_1(\Pi_a)$ Wilson lines, which are the $D=4$ scalars obtained upon dimensional reduction of the $D=7$ gauge field $A_M$ living on the D6-brane.

At low energies, we thus obtain a $D=4$ $\cn=1$ $U(N_a)$ gauge theory from the stack of $N_a$ D6-branes wrapping $\Pi_a$, with $b_1(\Pi_a)$ chiral multiplets transforming in the adjoint representation. This theory may also be equipped with a superpotential $W$ between the adjoint fields, generated either already at tree-level or by holomorphic open string world-sheet instantons \cite{kklm99,kklm00,vafa1,vafa2,vafa3}. This superpotential, however, may only freeze the $U(1)_a$ Abelian factor which is the trace of the $U(N_a)$ gauge group. Hence, it will not help when trying to get rid of adjoint fields of non-Abelian groups.

The most direct way of constructing intersecting D-brane models without adjoint matter multiplets is to consider SL 3-cycles $\Pi_a$ such that $b_1(\Pi_a) = 0, \ \forall a$. In practice, this will mean that $\Pi_a$ will have either the topology of $S^3$ or of any quotient of $S^3$ by a freely acting discrete group. This fact has been used in the literature in order to construct intersecting D-brane models with few adjoints fields, and where some D6-branes wrap 3-cycles with the topology of $S^3$ or $\mbb{RP}^3$. However, these models are either local \cite{AU02} (in the sense that the ${\bf CY_3}$ geometry is non-compact) or, in the case of the quintic, D6-branes wrapping simple $\mbb{RP}^3$ do not give rise to chiral models \cite{bbkl02}. One of the purposes of the present work is to show explicitly that global $\cn=1$ chiral models based on rigid intersecting D6-branes can indeed be constructed.

In constructing an intersecting D6-brane model, one considers several stacks of D6-branes, wrapped on different 3-cycles $\Pi_a$. In order to achieve an $\cn=1$ chiral model, we must require some intersection numbers $I_{ab}$ between these 3-cycles to be non-vanishing, as well as all the calibration phases $\phi_a$ to be all equal. If that is the case, we will recover a net number of $I_{ab}$ chiral multiplets in the bifundamental representation of $U(N_a) \times U(N_b)$, localized at the intersection points $\Pi_a \cap \Pi_b$. Of course, the scalar components of these chiral multiplets are new open string moduli of the theory, which cannot be made massive without breaking $\cn=1$ supersymmetry. We will come back to the issue of stabilizing those moduli in Section \ref{flux}, where we consider compactifications with background fluxes.

It turns out that requiring $\cn=1$ supersymmetry, and hence all the SL phases $\phi_a$ to be equal is incompatible with RR and NSNS tadpole cancellation, at least in this simple class of constructions. This obstruction can be overcomed by including some negative tension objects, such as O6-planes, in our construction. Indeed, O6-planes can be introduced by dividing our theory by $\Om \car$, where $\Om$ is the usual world-sheet parity and $\car$ an anti-holomorphic involution\footnote{To shorten the notation, we understand that $\car$ also includes the usual $(-1)^{F_L}$ factor.} which is a symmetry of $\cm_6$. The O6-planes will then wrap 3-cycles $\Pi_{O6}$ which are the fixed point set of $\car$, and are  automatically Special Lagrangian. 

Besides O6-planes, this orientifold quotient introduces several changes in the above picture. First, any stack of $N_a$ D6-branes wrapping a 3-cycle  invariant under $\car$ (i.e., $\Pi_a = \car \Pi_a$) will not yield a $U(N_a)$ gauge group, but rather an $SO(N_a)$ or $USp(N_a)$ gauge group. On the other hand, any stack $a$ wrapped on a cycle $\Pi_a$ not invariant under $\car$ will yield a $U(N_a)$ gauge group, but must be accompanied by its orientifold image $a'$, wrapped on $\Pi_{a'} = \car \Pi_a$. Second, D6-brane intersections will now contain chiral matter either in the bifundamental, symmetric or anti-symmetric representation of the gauge group \cite{bgkl00a}. Third, D6-branes which are non-BPS but stable and carry a torsion $\mbb{Z}_2$ RR charge may also appear. 

The RR tadpole conditions of this compactification are satisfied by imposing
\be
\sum_a N_a \left( [\Pi_a] + [\Pi_{a'}] \right) = 4\, [\Pi_{O6}]
\label{RRCY}
\ee
as well as the vanishing of the $\mbb{Z}_2$ torsion charges. $\cn=1$ supersymmetry imposes all the phases $\phi_a$ to be equal to the O6-plane phase $\phi_{O6}$, and this also guarantees the cancellation of NSNS tadpoles.

\subsection{Toroidal orientifolds}

Global constructions on background toroidal orbifolds yielding $\cn=1$ chiral vacua of the above class have indeed been achieved in the literature \cite{csu01a,bgo02,GH03,ho04}. Quite remarkably, some models admit gauge groups and chiral spectra rather close to either the Standard Model or proposed extensions of it.\footnote{In particular, the $\mbb{Z}_2 \ti \mbb{Z}_2$ background of \cite{csu01,csu01a} has been object of extensive model building research. See \cite{cps02,cp03a,cll04}.} These constructions are based on orientifolds of toroidal orbifolds of the form $(\T^2 \times \T^2 \times \T^2)/\Gamma$ \cite{bgk99,fhs00} and, as we will now review, they all employ non-rigid 3-cycles.\footnote{To our knowledge, the only D-brane models in the literature where rigid cycles implicitly appear are the brane susy-breaking models in \cite{aaads99} and some shifted orientifold models in \cite{aads99,lp03}.} We have summarized such orbifold backgrounds and their Hodge numbers in table \ref{orbi}.

\begin{table}[htb]
\renewcommand{\arraystretch}{1.25}
\begin{center}
\begin{tabular}{|c||c|c|c|c|c|c|c|c|c|}
\hline
$\Gamma$ &  $\mbb{Z}_3$ & $\mbb{Z}_4$ &  $\mbb{Z}_6$ &  $\mbb{Z}_6'$ & $\mbb{Z}_2  \times \mbb{Z}_2$ & $\mbb{Z}_2  \times \mbb{Z}_2^\prime$  & $\mbb{Z}_2  \times \mbb{Z}_4$ & $\mbb{Z}_3  \times \mbb{Z}_3$ & $\mbb{Z}_3  \times \mbb{Z}_6$ \\
\hline \hline
$h_{11}^{\rm unt}$ & 9 & 5 & 5 & 3 & 3 & 3 & 3 & 3 & 3 \\
\hline
$h_{11}^{\rm tw}$ & 27 & 26 & 24 & 32 & 48 & 0 & 58 & 81 & 70 \\
\hline
\hline
$h_{21}^{\rm unt}$ & 0 & 1 & 0 & 1 & 3 & 3 & 1 & 0 & 0 \\
\hline
$h_{21}^{\rm tw}$ & 0 & 6 & 5 & 10 & 0 & 48 & 0 & 0 & 1 \\
\hline
\end{tabular}
\caption{\small Orbifold backgrounds of the form $(\T^2 \times \T^2 \times \T^2)/\Gamma$ yielding $\cn=1$ orientifold vacua with D6-branes at angles. For a classification of general $\T^6/\mbb{Z}_N$ $\cn=1$ orientifolds see \cite{bcs04}.}
\label{orbi}
\end{center}
\end{table}

In these orbifold backgrounds, the number of independent 3-cycles that D6-branes can wrap is given by $b_3\, (\T^6/\G) = 2 + 2h_{21}^{\rm unt} + 2h_{21}^{\rm tw}$. Of these, $2 + 2h_{21}^{\rm unt}$ 3-cycles can be thought as inherited from the homology of the covering space $\T^6$. These are usually called 
 {\em bulk} 3-cycles, and are of the form
\be
\Pi_a^B = \sum_{g \in \G} \Th_g\cdot \Pi_a^{\T^6},
\label{bulk}
\ee
where $\Pi_a^{\T^6}$ is a 3-cycle on the covering space and $\Th_g$ is the geometrical action of the group element $g$. In addition, $2h_{21}^{\rm tw}$ {\em twisted} 3-cycles will be present in the orbifolded geometry, this time arising from fixed loci of the orbifold action. 

Building $\cn=1$ chiral vacua from these orientifold backgrounds is a non-trivial task. Actually, not all the orientifolds constructed from table \ref{orbi} admit $\cn=1$ chiral vacua. For instance, both $\mbb{Z}_3$ and $\mbb{Z}_3 \times \mbb{Z}_3$ have $b_3 = 2$, and hence only two SL 3-cycles independent in homology. One of these 3-cycles is calibrated by $\preal (e^{i\pi/2} \Om_3)$, and the other by $\preal (\Om_3)$, and hence have different phases. Since in order to preserve $\cn=1$ supersymmetry all the D6-branes have to be calibrated by the same phase as the O6-plane, we find that $[\Pi_a] = [\Pi_{a'}] = [\Pi_{O6}]$, and no chiral spectrum arises.

The first orbifold employed in constructing $\cn=1$ chiral vacua of intersecting D6-branes was $\mbb{Z}_2 \times \mbb{Z}_2$ \cite{csu01,csu01a}, usually referred to as $\mbb{Z}_2 \times \mbb{Z}_2$ without vector structure. As can be seen from table \ref{orbi} this orbifold has eight independent 3-cycles, all of them inherited from the covering space $\T^6$. When constructing a bulk 3-cycle as in (\ref{bulk}), one finds that under the action of $\mbb{Z}_2 \times \mbb{Z}_2$ a 3-cycle on $\T^6$ is always mapped to a parallel 3-cycle identical in homology. Hence we obtain that $[\Pi_a^B] = 4 [\Pi_a^{\T^6}]$. The moduli space of such bulk D-branes has been investigated, from a T-dual point of view, in \cite{d98}. From a similar analysis one concludes that one bulk D6-brane will contain an $\cn=4$ $U(1)$ Super Yang-Mills theory in its world-volume,\footnote{Broken to $\cn=1$ by $g_s$ and $\alpha'$ corrections. That is, by taking into account the coupling of the massless open string states to the closed string twisted fields as well as to the massive open string states arising from the D6-brane $a$ and its orientifold images.} and that the three adjoint fields of such a theory parametrize the moduli space of the D6-brane, which is nothing but $\T^6/(\mbb{Z}_2 \times \mbb{Z}_2)$. Notice that this open string moduli space is singular, and that the singularities correspond to the D6-brane going through $\mbb{Z}_2 \times \mbb{Z}_2$ fixed points on each of the three $\T^2$'s as, e.g., when crossing the origin of $\T^2 \times \T^2 \times \T^2$. When this happens, the D6-brane gauge symmetry will be enhanced to $U(2)$, whereas supersymmetry will be broken from $\cn=4$ to $\cn=1$ by means of a superpotential for the adjoint fields $W = \prod_{i=1}^3 \Phi^i$. 

Now, when understanding the space of SL 3-cycles of an orbifold background, it is important to construct a self-dual/unimodular lattice of $H_3(\T^6/\G,\mbb{Z})$. It turns out that the bulk 3-cycles $\Pi^B$ above do not expand such a basis, and that we must instead consider {\em fractional} 3-cycles of the form $\Pi_a^f = \half \Pi_a^B$. Such a fractional D6-brane crossing  $\mbb{Z}_2 \times \mbb{Z}_2$ fixed points will yield an $\cn=1$ $U(1)$ gauge theory (in contrast to $U(2)$ of a bulk D6-brane) with three chiral multiplets in the adjoint and the superpotential $W = \prod_{i=1}^3 \Phi^i$. Notice that, due to the presence of $W$ the moduli space of $\Pi_a^f$ is now of complex dimension one. Indeed, we can still move the D-brane away from the fixed points, but now in only one of the three $\T^2$'s. 

In general, in order to construct an $\cn=1$ chiral model in $\mbb{Z}_2 \times \mbb{Z}_2$, we will combine stacks of $N_a$ fractional and bulk D6-branes, as well as D6-branes parallel to the O6-planes. The latter do not yield $U(N)$ but rather $USp(2N)$ gauge groups, as well as chiral multiplets in the anti-symmetric
representation. In any case, the above facts generalize to configurations with gauge groups $\prod_a U(N_a) \times \prod_b USp(2N_b)$,\footnote{For a detailed analysis of this case and its expected matching with field theory Higgsing see \cite{clll04}.} and as a general result we find that any $\cn=1$ chiral compactification yields plenty of open string moduli corresponding to D6-branes translations, and in particular three massless adjoints for each $U(N)$ factor.

Nevertheless, the $\T^6/(\mbb{Z}_2 \times \mbb{Z}_2)$ orbifold is extremely simple in the sense that every 3-cycle is somehow inherited from $\T^6$. In general, both bulk and twisted 3-cycles will contribute to $H_3(\T^6/\G,\mbb{Z})$, and the elements of the integral homology basis will be given by linear combinations of both of them. Notice that a 3-cycle wrapping bulk and twisted cycles at the same time will be stuck at the orbifold fixed points where the twisted 3-cycles arise, and hence the adjoint field corresponding to transverse translations of the D6-brane will not be present. Our purpose in this work is to show how such rigid 3-cycles without adjoints can arise from an orientifold background, and to construct explicit $\cn=1$ chiral models by means of intersecting D6-branes wrapping rigid cycles. In order to accomplish our task, in the next section we will focus on the simplest orbifold background containing rigid 3-cycles, namely $\mbb{Z}_2\times \mbb{Z}_2'$ of table \ref{orbi}, also known as $\mbb{Z}_2\times \mbb{Z}_2$ with discrete torsion.

One may naively think that these rigid cycles arise whenever one considers an orientifold with twisted 3-cycles, and that any fractional D6-brane wrapping a twisted cycle will indeed be rigid. That would mean that the $\cn=1$ chiral models in \cite{bgo02,ho04} are free of adjoint matter fields, which is not the case. On the contrary, it happens that for all $\mbb{Z}_N$  orbifolds with $N > 2$ there exist a second source of adjoint scalars. Indeed, two different orbifold images of the same 3-cycle, $\Th_g \Pi_a$ and $\Th_h \Pi_a$ with $g\neq h$, will in general intersect non-trivially, and each of these intersections yields a chiral multiplet in the adjoint \cite{bklo01}. Therefore, even though it might seem that orbifolds with twisted cycles help to reduce the number of adjoints, these may reappear through the back-door. This has been shown explicitly for fractional D6-branes in the $\T^6/\mbb{Z}_4$ orbifold in \cite{bgo02}.

\subsection{Other constructions}

Another recently discussed class of intersecting D-brane models are Gepner model orientifolds \cite{aaln03,RB03a,bhhw04,bw04,dhs04,bw04a,aaj04,dhs04a}. Since Gepner models describe exact solutions of the non-linear sigma model on certain Calabi-Yau manifolds deep inside the K\"ahler moduli space, they provide some insights on 3-cycles on Calabi-Yau manifolds there. Even though one is loosing  a direct geometric interpretation and F-terms are generally generated as one deforms the K\"ahler parameters, the general formulae for the boundary and crosscap states allow to read off the massless sector in each open string sector. The search carried out in \cite{dhs04a} revealed that some of these branes also do not have additional adjoints and should thus be considered as wrapping rigid cycles. 

There is also some recent work on the so-called shifted $\mbb{Z}_2\times \mbb{Z}_2$ orientifolds \cite{aads00shift,GP02,LG03,GP03}. These constructions involve D-branes on top of the orientifold planes and with non-trivial magnetic fluxes on them. Looking carefully at the models discussed in \cite{LG03}, one realizes that they  contain completely rigid branes (named ${\cal N}=1$ supersymmetric branes in the above reference). The geometric construction of these rigid cycles is somewhat analogous to what we will present in this paper. More precisely, due to the additional shifts in the $\mbb{Z}_2$ actions, there appear twisted 3-cycles in each $\mbb{Z}_2$ twisted sector. Then, by considering certain fractional D-branes running through fixed point of all $\mbb{Z}_2$ symmetries completely freezes the positions of the branes, yielding a rigid D-brane.

\section{Intersecting rigid branes on $\mbb{Z}_2\times \mbb{Z}_2'$}
\label{intersecting}

In this section we develop the necessary techniques to build $\cn=1$ chiral vacua based on rigid intersecting D6-branes. Motivated by the discussion of the previous section, we will focus on the $\T^6/(\mbb{Z}_2\times \mbb{Z}_2')$ orientifold background of table \ref{orbi}. Notice that this background is related by T-duality to one of the {\em brane supersymmetry breaking} constructions of \cite{aaads99}, and hence one could naively think that no $\cn=1$ vacua could be found. However, as shown explicitly in \cite{ms04a}, such vacua do exist. In the next section we will generalize the simple $\cn=1$ example presented in \cite{ms04a}, in order to construct $\cn=1$ chiral vacua involving rigid D6-branes.

\subsection{The orbifold background}

Let us consider type IIA string theory compactified on the orbifold background $\T^6/(\mbb{Z}_2\times \mbb{Z}_2)$, where the $\mbb{Z}_2$ generators act as
\be
\Theta:\cases{z_1\to -z_1 \cr
              z_2\to -z_2 \cr
              z_3\to z_3 \cr }\quad\quad\quad\quad
\Theta':\cases{ z_1\to z_1 \cr
                z_2\to -z_2 \cr
                 z_3\to -z_3 \cr }
\label{orbiaction}
\ee 
on the three complex coordinates of $\T^6 = \T^2 \times \T^2 \times \T^2$. As explained in \cite{vafa86}, this is not enough to specify the string background completely, as we have two inequivalent choices of discrete torsion relating the two $\mbb{Z}_2$ orbifold actions. One of these choices corresponds to the Hodge numbers $(h_{11}, h_{21}) = (3,51)$ and the other one to $(h_{11}, h_{21}) = (51,3)$. These backgrounds are related by mirror symmetry, and will be referred here as $\mbb{Z}_2 \times \mbb{Z}_2$ orbifolds with and without discrete torsion, respectively.

In the present paper we will be mainly interested in the $\mbb{Z}_2 \times \mbb{Z}_2$ orbifold with discrete torsion, denoted in the following as $\mbb{Z}_2 \times \mbb{Z}_2'$. The reason is that the twisted homology of this orbifold contains collapsed 3-cycles, and these will allow us to construct rigid 3-cycles that D6-branes can wrap. More precisely, $\T^6/(\mbb{Z}_2\times \mbb{Z}_2')$ contains eight untwisted and non-rigid 3-cycles, all of them inherited from the covering space $(\T^2)^3$. The rest of the homology group $H_3$ is made of $3 \times 32$ twisted 3-cycles, giving a total of $b_3 =104$. 

Indeed, the twisted homology of a $\mbb{Z}_2 \times \mbb{Z}_2$ orbifold can be understood as follows. If we consider the six-torus quotiented by $\Theta$, we recover a ${\bf K3} \times \T^2$ geometry, where ${\bf K3}$ is in its orbifold limit $\T^4/\mbb{Z}_2$. Such $\T^4/\mbb{Z}_2$ orbifold contains 16 fixed points, which after blowing up give rise to the 16 additional 2-cycles with the topology of ${\bf P}^1 \simeq S^2$. The twisted homology of ${\bf K3} \times \T^2$ is then given by tensoring these 16 twisted 2-cycles with the homology of $\T^2$. We find $h_{11}^{\rm tw} = h_{22}^{\rm tw} = 16$ and $h_{21}^{\rm tw} = h_{12}^{\rm tw} = 16$. We now have to take into account the action of $\Theta'$ on this twisted homology space and, in particular, on the collapsed ${\bf P}^1$'s of ${\bf K3}$. This action is of the form $\Theta': {\bf P}^1 \mapsto \eta\,  {\bf P}^1$, where $\eta = \pm 1$ represents the choice of discrete torsion. Hence, in the $\mbb{Z}_2 \times \mbb{Z}_2$ orbifold without discrete torsion ($\eta = +1$) the invariant twisted cycles are $h_{11}^{\rm tw} = h_{22}^{\rm tw} = 16$, whereas in the orbifold with discrete torsion ($\eta = -1$) these are $h_{21}^{\rm tw} = h_{12}^{\rm tw} = 16$. We thus see that in the latter case we recover 32 collapsed 3-cycles per twisted sector, up to a total of 96 twisted 3-cycles with the topology of $S^2 \times S^1$.

Intuitively, we see that a D6-brane wrapping collapsed 3-cycles in each of the three twisted sectors is stuck at some particular position on the covering space $\T^2 \times \T^2 \times \T^2$. In that way, we recover D6-branes wrapping rigid 3-cycles. We will perform a more detailed analysis of these rigid cycles in the following subsections, making use of the formalism developed in \cite{bbkl02} to describe 3-cycles in toroidal orbifolds.

\subsection{Rigid 3-cycles}

\begin{figure}
\begin{center}
\epsfbox{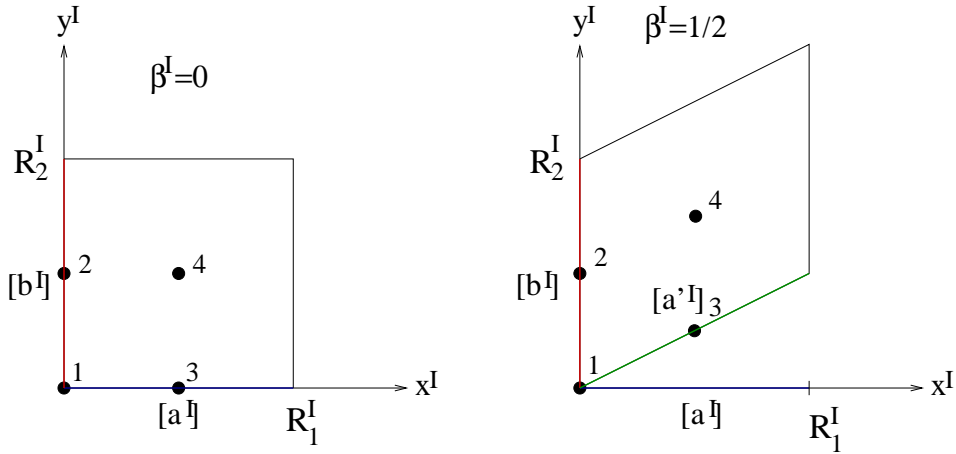}
\caption{Choices of complex structures on the $\T^2$ factors and $\mbb{Z}_2$ fixed points}
\label{fct2}
\end{center}
\end{figure}

In order to describe the rigid 3-cycles in $\T^6/(\mbb{Z}_2 \times \mbb{Z}_2')$, let us first consider the covering space $(\T^2)_1 \times (\T^2)_2 \times (\T^2)_3$, and introduce complex coordinates of the form $z^I = x^I + iy^I$ on each of the $(\T^2)_I$ factors, $I = 1,2,3$. As shown in figure \ref{fct2}, on each $\T^2$ there exist two different choices of the complex structure which are compatible with the anti-holomorphic involution $\car:z^I \to \overline z^I$ \cite{bkl00}.\footnote{At this level of the construction we could consider an arbitrary complex structure. When constructing $\cn=1$ models, however, the anti-holomorphic involution $\car$ will play an essential role (see subsection \ref{ori}). Hence, in the following we restrict our attention to complex structures that admit $\car$ as a symmetry.}
%
%%%%%%%%%%%%%%%%%%%%%%%%%%%%%%%%%%%%%
%%%%%%%%%%%%%%%%%%%%%%%%%%%%%%%%%%%%%
%
Next, we introduce factorizable D6-branes at angles, which are specified by the wrapping numbers $(n^I, m^I)$ along the fundamental 1-cycles $[a^I]$ and $[b^I]$, respectively $[a'^I]$ and $[b^I]$, on each $\T^2$. As usual, it is convenient to express the set of 1-cycles in terms of $[a^I]$ and $[b^I]$, by simply writing $[a'^I]=[a^I]+{1\over 2}[b^I]$. A factorizable 3-cycle on $(\T^2)_1 \times (\T^2)_2 \times (\T^2)_3$ is defined as the product of three 1-cycles 
\be \label{torcycle}
\Pi_a=\bigotimes_{I=1}^3 \left( n^I_a\ [a^I] + \widetilde m^I_a\ [b^I] \right)
\ee
with $\widetilde m^I_a = m^I_a+\beta^I\,  n^I_a$ (see fig.\ref{fct2} for the definition of $\b^I$).

These 3-cycles of $(\T^2)^3$ will be inherited by the orbifold quotient. However, in order to deal with 3-cycles on orbifold spaces we have to carefully distinguish between 3-cycles on the covering space and 3-cycles on the actual orbifold \cite{bbkl02}. In the particular case at hand, under the action of $\mbb{Z}_2\times \mbb{Z}_2'$ a factorizable 3-cycle on $\T^6$ has 3 images, all of them with the same wrapping numbers as the initial 3-cycle. Therefore, a 3-cycle in the {\it bulk} of the orbifold space can be identified with $[\Pi_a^B] = 4\, [\Pi_a^{\T^6}]$. Computing the intersection number we get
\be \label{prod}
[\Pi_a^B] \cdot [\Pi_b^B] = 
4\, [\Pi_a^{\T^6}] \cdot [\Pi_b^{\T^6}] =
4 \prod_{I=1}^3 (n^I_a\, \widetilde m^I_b - \widetilde m^I_a\, n^I_b) = 
4\prod_{I=1}^3 (n^I_a\, m^I_b - m^I_a\, n^I_b),
\ee
where we have identified the intersection points related by the $\mbb{Z}_2\times \mbb{Z}_2'$ action.

In addition to these untwisted cycles we have 32 independent collapsed 3-cycles for each of the three twisted sectors, $\Theta$, $\Theta'$ and $\Theta\Theta'$. Let us first consider the $\Theta$ twisted sector. We denote the 16 fixed points on $(\T^2)_1 \times (\T^2)_2/\mbb{Z}_2$ by $[e^\Theta_{ij}]$, with $i,j\in\{1,2,3,4\}$ (see fig. \ref{fct2}). After blowing up the $\mbb{Z}_2$ orbifold singularities, these become two-cycles with the topology of $S^2$. Given our choice of discrete torsion, these two-cycles are combined with a one-cycle of $(\T^2)_3$, $n^3 [a^3] + \widetilde m^3\ [b^3]$, in order to form a 3-cycle in the $\Theta$-twisted sector. Let us denote a basis of such twisted 3-cycles as
\be
[\alpha^\Theta_{ij,\,n}] = 2\, [e^\Theta_{ij}]\otimes  [a^3], \quad\quad
[\alpha^\Theta_{ij,\,m}] = 2\, [e^\Theta_{ij}]\otimes [b^3], 
\ee 
where the extra factor of two is due to the action of $\Th'$ on the twisted 3-cycles 
in general position on the third $\T^2$ factor. Analogously, we define the basic twisted 3-cycles in the $\Theta'$ and $\Theta\Theta'$ twisted sectors as 
\be
\begin{array}{lcl} \vspace*{.2cm}
\quad [\alpha^{\Theta'}_{ij,\,n}] = 2\, [e^{\Theta'}_{ij}]\otimes [a^1] & &
[\alpha^{\Theta'}_{ij,\,m}] = 2\, [e^{\Theta'}_{ij}]\otimes [b^1] \\
\quad [\alpha^{\Theta\Theta'}_{ij,\,n}] = 2\, [e^{\Theta\Theta'}_{ij}]\otimes [a^2]
& & 
[\alpha^{\Theta\Theta'}_{ij,\,m}] = 2\, [e^{\Theta\Theta'}_{ij}] \otimes [b^2].
\end{array}
\ee
The intersection number between a pair of such cycles is easy to compute knowing that the collapsed $\P^1$'s of $\K3$ have self-intersection number $[e_{ij}] \cdot [e_{kl}] = -2 \d_{ik} \d_{jl}$, and that two $\P^1$'s of different twisted sectors do not intersect. Given the 3-cycles $[\Pi^g_{ij,\,a}]= n_a^{I_g} [\alpha_{ij,\,n}] + \widetilde m_a^{I_g} [\alpha_{ij,\,m}]$ and $[\Pi^h_{kl,\,b}]= n_b^{I_h}[\alpha_{kl,\,n}] + \widetilde m_b^{I_h} [\alpha_{kl,\,m}]$, with $g,h = \Th, \Th', \Th\Th'$, we find
\be 
\label{inttwist} 
[\Pi^g_{ij,\,a}] \cdot [\Pi^h_{kl,\,b}]\,
=\, 4\,\sigma\,  \d_{ik} \d_{jl} \d^{gh} \,
(n_a^{I_g}\, \widetilde m_b^{I_g} - \widetilde m_a^{I_g}\, n_b^{I_g})\, 
=\, 4\, \sigma\, \d_{ik} \d_{jl} \d^{gh} \,
(n_a^{I_g}\, m_b^{I_g} - m_a^{I_g}\, n_b^{I_g})
\ee
where we have again identified intersection points under the orbifold action. In this notation, for the twisted sectors $g = \Theta, \Theta',\Theta\Theta'$ one has $I_g = 3,1,2$, respectively. Here $\sig$ is a sign that can be fixed by noticing that the conventions (\ref{torcycle}) and (\ref{prod}) for the intersection product imply that $[a^I] \cdot [b^J] = - \d_{IJ}$, and hence $\sig = +1$ in all three twisted sectors. Indeed, this will turn out to be the correct choice in the sense that it guarantees integer intersection numbers between fractional rigid D6-branes, as well as the one consistent with the conformal field theory computations performed in \cite{aaads99} which involve simple D-branes on top of the orientifold planes.

\begin{figure}
\epsfxsize=5.75in
\begin{center}
\epsfbox{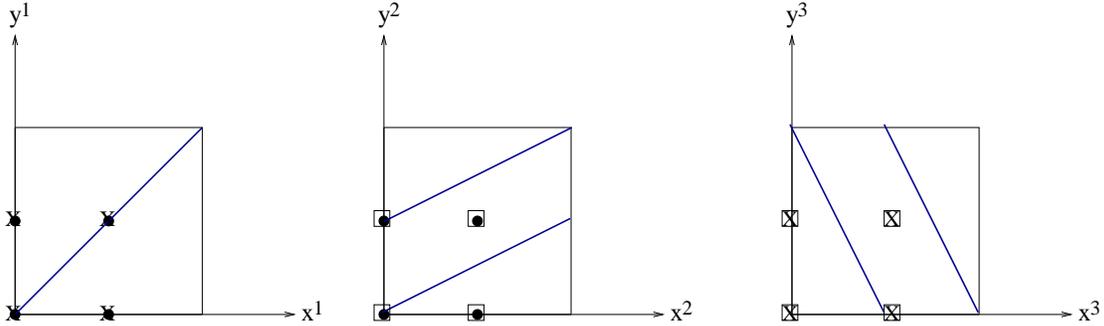}
\caption{Fractional brane running through 4 fixed points for each twisted sector. Fixed points are denoted by dots in the $\Theta$, by squares in the $\Theta'$ and by crosses in the $\Theta\Theta'$ twisted sector. \label{fixedpoints}}
\end{center}
\end{figure}

Equipped with the above description of the untwisted and twisted sector 3-cycles, it is now clear how to build rigid D6-branes in this framework. Namely, we will consider fractional D6-branes which are wrapping Special Lagrangian 3-cycles  and are  charged under all three different twisted sectors of the orbifold. In order to construct such D-branes, let us start with a factorizable 3-cycle, described by three pairs of wrapping numbers  $(n_a^I, m_a^I)$. A fractional D6-brane should be invariant under the orbifold action, and hence it must run through four fixed points for each twisted sector, as illustrated in fig. \ref{fixedpoints}.
%
%%%%%%%%%%%%%%%%%%%%%%%%%%%%%%
%%%%%%%%%%%%%%%%%%%%%%%%%%%%%%%
%
Let us denote this set of four fixed points on the twisted sector $g$ as $S_g^a$. Then the entire 3-cycle that such a fractional D-brane is wrapping is of the form
\be 
\label{rigid}
\Pi^F_a\, =\, {1\over 4}\, \Pi^B_a +
{1\over 4} \left( \sum_{i,j \in S_\Theta^a} \epsilon^\Theta_{a,ij}\,  
\Pi^\Theta_{ij,\,a} \right)+
 {1\over 4} \left( \sum_{j,k\in S_{\Theta'}^a} \epsilon^{\Theta'}_{a,jk}\, 
\Pi^{\Theta'}_{jk,\,a} \right)+
 {1\over 4}\left( \sum_{i,k\in S_{\Theta\Theta'}^a} 
\epsilon^{\Theta\Theta'}_{a,ik}\, \Pi^{\Theta\Theta'}_{ik,\,a} \right)
\ee
where the $1/4$ factor indicates that one needs four such fractional branes in order to get a bulk brane. Here the signs $\epsilon^\Theta_{a,ij},\,\epsilon^{\Theta'}_{a,jk},\,\epsilon^{\Theta\Theta'}_{a,ik}\,=\,\pm 1$ define the charge of the fractional brane $a$ with respect to the massless fields living at the various fixed points. Geometrically, these numbers indicate the two possible orientations with which the brane can wrap around the blown up $\P^1$. Both the sets of fixed points $S_g$  and the signs $\eps^g$ are not arbitrary, but are constrained by several consistency conditions which we proceed to analyze in some detail.

\subsection{Discrete positions and Wilson lines}

A fractional D-brane of the form (\ref{rigid}) is specified by several sets of topological data. Namely, the untwisted wrapping numbers $(n^i,m^i)$, for $i=1,2,3$, the sets of fixed point $S_g$ and the signs $\eps^g$, for $g = \Th, \Th', \Th\Th'$. In the following we show that, given the wrapping numbers $(n^i,m^i)$, there are exactly $8 \times 8 = 64$ inequivalent choices of fractional D6-branes that one can consider. Indeed, in a T-dual picture, this freedom exactly matches with the 64 different positions that a fractional D3-brane can be located, or the different Wilson lines that can be turned on for a fractional D9-brane. As we will see, in the  present case the choice of $S_g$ is related to the  discrete positions that a fractional D-brane with wrapping numbers $(n^i,m^i)$ can have, whereas the signs $\eps$ in (\ref{rigid}) are related to the discrete Wilson lines that can be turned on them.

From figure \ref{fixedpoints} it is easy to see that the sets $S_g$ are specified by the position of the fractional D6-brane, and that there are only a finite number of such choices. In general, given the bulk wrapping numbers $(n^i,m^i)$ of a fractional D6-brane, one may wonder through which fixed points it can  run. Let us first discuss this problem for a single square $\T^2$ factor. More precisely, we consider a (fractional) 1-cycle with wrapping numbers $(n,m)$ on $\T^2/\mbb{Z}_2$. Such 1-cycle runs through a pair of fixed points, so in principle there are six possibilities given by $\{1,2\}$, $\{1,3\}$, $\{1,4\}$, $\{3,4\}$, $\{2,4\}$ or $\{2,3\}$. However, it is clear that a 1-cycle $(n,m)$ which intersects $\{1,2\}$ may go through $\{3,4\}$ by a simple transverse translation. On the other hand, it would never intersect, say, $\{1,3\}$. It is in fact easy to characterize the sets of pairs of fixed points in terms of the wrapping numbers $(n,m)$, the result being shown in table \ref{fixed}.

\begin{table}[htb]
\renewcommand{\arraystretch}{1.25}
\begin{center}
\begin{tabular}{|c|c|c|}
\hline
$(n,m)$ & Fixed points \\
\hline \hline
(odd, odd) & $\{1,4\}$ or $\{2,3\}$\\
(odd, even) & $\{1,3\}$ or $\{2,4\}$\\
(even, odd) & $\{1,2\}$ or $\{3,4\}$\\
\hline
\end{tabular}
\caption{\small Fixed points of a 1-cycle on a $\T^2/\mbb{Z}_2$ in terms of its wrapping numbers.}
\label{fixed}
\end{center}
\end{table}

It is now straightforward  to see how this result generalizes for fractional 3-cycles on $\T^6/(\mbb{Z}_2 \times \mbb{Z}_2)$. We are now interested in knowing which are the fixed points $(i,j) \in S_g^a$ given the bulk wrapping numbers $\Pi_a^B = [(n_a^1,m_a^1)] \otimes  [(n_a^2,m_a^2)] \otimes [(n_a^3,m_a^3)]$. In general, $S_g^a$ is a subset of $2 \times 2$ elements inside $\{1,2,3,4\} \times \{1,2,3,4\}$. Let us, for instance, consider $S_\Th^a$, which can be computed by looking at $(n_a^1,m_a^1)$, $(n_a^2,m_a^2)$  and finding their fixed point set by means of table \ref{fixed}. The index $i$ will take two different values, determined by $(n_a^1,m_a^1)$ and one choice in table \ref{fixed}. For instance, if $(n_a^1,m_a^1) = ({\rm odd}, {\rm odd})$, $i$ will run through either $\{1,4\}$ or $\{2,3\}$, and this basically amounts to specify the position of the rigid cycle in the first $\T^2$. Something similar will happen for $j$ with respect to $(n_a^2,m_a^2)$. In total, for any given bulk 3-cycle $\Pi^B_a$ we always have 8 different choices for placing a fractional D6-brane $\Pi_a^F$ and, correspondingly, 8 different choices for specifying $S_\Th^a$, $S_{\Th'}^a$ and $S_{\Th\Th'}^a$. 

Notice, however, that two fractional D6-branes $\Pi_{a1}^F$ and $\Pi_{a2}^F$ with the same bulk component $\Pi^B_{a}$ and different choice of $S_g^a$ do {\em not} correspond to the same homology class of 3-cycles, since their collapsed 3-cycles are different. In this sense, each of the 8 different rigid positions associated to $\Pi^B_{a}$ in the discussion above cannot be thought as a discrete moduli space of a rigid D6-brane. On the contrary, each discrete position correspond to a different fractional D6-brane.

Let us now turn to the freedom associated to the signs $\eps_{ij}^g$, which are related to the collapsed cycles that the D6-branes wrap. In principle it seems that we have $2^{16}$ different choices for these signs. However, there are some consistency constraints that need to be imposed on $\eps_{ij}^g$. Indeed, as explained in \cite{Sen98}, the $\eps$'s are related to the discrete Wilson lines along the brane, so they have to satisfy
\be
\label{bedh}
\sum_{i,j\in S_g^a} \epsilon^g_{a,ij}= 0\ {\rm mod}\ 4
\ee
and similarly for the other two twisted sectors.\footnote{The other choices of signs correspond to turning on not only constant Wilson lines but constant magnetic  fluxes on the brane, something we do not consider in this paper.} Moreover, there exist two  more important conditions on these signs factors. Indeed, the signs in different twisted sectors are not completely unrelated and need to fulfill the conditions
\be
\begin{array}{rcl} \vspace{.2cm}
\epsilon^\Theta_{a,ij}\,  \epsilon^{\Theta'}_{a,jk}\,
\epsilon^{\Theta\Theta'}_{a,ik}&=&1, \nonumber \\
\epsilon^\Theta_{a,ij}\,  \epsilon^{\Theta'}_{a,jk}&=&{\rm const.} \quad \forall j
\end{array}
\label{bedi}
\ee
where no summation over $i,j,k$ is performed. Such conditions are required for factorization of one loop amplitudes and are to be expected as only two of the three $\mbb{Z}_2$ actions, $\Theta,\Theta',\Theta\Theta'$, are independent. The first condition guarantees, in particular, that one indeed needs four such fractional branes to build a pure bulk brane, whereas the second condition is necessary for the first one to be well defined.

In addition, there is some redundancy in the definition of $\Pi_a^F$ in (\ref{rigid}). Indeed, if we allow $(n_a^I, m_a^I) \in \mbb{Z}^2$, then we should identify two 3-cycles which are related under
\bea \nonumber
(n_a^1, m_a^1) \otimes (n_a^2, m_a^2) \otimes (n_a^3, m_a^3) & \mapsto &
(-n_a^1, -m_a^1) \otimes (-n_a^2, -m_a^2) \otimes (n_a^3, m_a^3) \\
\label{redundancy}
\eps_{a,ij}^{\Th'} & \mapsto & - \eps_{a,ij}^{\Th'} \\ \nonumber
\eps_{a,ij}^{\Th\Th'} & \mapsto & - \eps_{a,ij}^{\Th\Th'}
\eea
and the same for permutations of the three $\T^2$'s. Notice that these identifications are compatible with the conditions (\ref{bedi}). 

Let us now take all this into account, in order to find the most general set of $\eps_{ij}^g$ phases. In general, the set of twisted charges of a fractional D6-brane on $\Pi_a^F$ is of the form
\be
\begin{array}{rcl}
S_{\Th} & = & \left\{ \{i_1, i_2\} \times \{j_1, j_2\} \right\} \\
S_{\Th'} & = & \left\{ \{j_1 j_2\} \times \{k_1 k_2\} \right\} \\
S_{\Th\Th'} & = & \left\{ \{k_1 k_2\} \times \{i_1 i_2\} \right\}
\end{array}
\label{twistedsets}
\ee
where $i_\a, j_\a, k_\a$, $\a =1,2$ represent fixed point coordinates in the first, second and third $\T^2$ factors, respectively.

Now, taking the redundancy (\ref{redundancy}) into account, one can fix one of the four signs $\eps_{ij}^g$ for each $S_g$. Let us, for instance, choose $\eps_{i_1j_1}^{\Th} = \eps_{j_1k_1}^{\Th'} = \eps_{k_1i_1}^{\Th\Th'} = +1$. Then, by imposing (\ref{bedh}) we see that the signs $\eps_{ij}^g$ on each set $S_g$ have to be of the form $\{1,\i, \i', \i\i' \}$, with $\i,\i' = \pm 1$. By further imposing eqs.(\ref{bedi}) we arrive to the general solution shown in table \ref{signs}.

\begin{table}[htb]
\renewcommand{\arraystretch}{1.5}
\begin{center}
\begin{tabular}{|c||cc|cc|cc|}
\hline
\hline
& $i_1$ & $i_2$ & $j_1$ & $j_2$ & $k_1$ & $k_2$ \\
\hline \hline
$i_1$ & & & $+1$ & $\i'$ & &  \\
$i_2$ & & & $\i$ & $\i\i'$ & & \\
\hline
$j_1$ & & & & & $+1$ & $\i''$  \\
$j_2$ & & & & & $\i'$ & $\i'\i''$ \\
\hline
$k_1$ & $+1$ & $\i$ & & & &  \\
$k_2$ & $\i''$ & $\i\i''$ & & & & \\
\hline
\end{tabular}
\caption{\small Inequivalent solutions of the constraints (\ref{bedh},\ref{bedi}). Here $\i, \i', \i'' = \pm 1$. In total we recover 8 inequivalent choices for the signs 
$\eps_{ij}^g$.}
\label{signs}
\end{center}
\end{table}

Notice that in total we obtain $2^3 = 8$ inequivalent choices of signs.  This was somehow to be expected, since these inequivalent choices should correspond to the choice of discrete Wilson lines, along the fractional D6-brane. Indeed, table \ref{signs} suggests a simple interpretations of the signs $\i, \i', \i''$ in terms of discrete Wilson lines. Namely, a translation from the point $i_1$ to the point $i_2$ in the first $\T^2$ is associated to a phase given by $\i$. The same interpretation can be driven for $\i'$, $\i''$ and the distances $j_1j_2$, $k_1k_2$, respectively. 

Finally notice that, after fixing both $S_g$ and $\eps_{a,ij}^g$ for a fractional D6-brane, there are four inequivalent choices of wrapping numbers $(n_a^i, m_a^i)$ which correspond to the same bulk 3-cycle $\Pi_a^B$ but different fractional 3-cycle $\Pi_a^F$. These are given by
\be
\begin{array}{ccccc}
\quad [(n_a^1,m_a^1)] & \otimes & [(n_a^2,m_a^2)] & \otimes & [(n_a^3,m_a^3)] \\
\quad [(-n_a^1,-m_a^1)] & \otimes & [(-n_a^2,-m_a^2)] & \otimes & [(n_a^3,m_a^3)] \\
\quad [(n_a^1,m_a^1)] & \otimes & [(-n_a^2,-m_a^2)] & \otimes & [(-n_a^3,-m_a^3)] \\
\quad [(-n_a^1,-m_a^1)] & \otimes & [(n_a^2,m_a^2)] & \otimes & [(-n_a^3,-m_a^3)] \\
\end{array}
\label{ineqbulk}
\ee
and are to be interpreted as the four different $\mbb{Z}_2 \times \mbb{Z}_2$ twisted charges that such a fractional D6-brane can have. Indeed, one needs these four fractional D6-branes in order to build the regular representation of $\mbb{Z}_2 \times \mbb{Z}_2$ and, eventually, a bulk D6-brane.

\subsection{The orientifold projection}\label{ori}

In general, in order to build $\cn=1$ models based on D6-branes on ${\bf CY_3}$ backgrounds we need to introduce O6-planes in our construction. As usual in type IIA intersecting D6-brane models, in the $\mbb{Z}_2\times \mbb{Z}_2'$ orbifold we achieve this by further modding out the theory by $\Om \car$, where the anti-holomorphic involution $\car:z_I\to \overline{z}_I$ is simply complex conjugation. In our case this introduces four types of orientifold $O6$-planes, which are located at the fixed point loci of $\Omega \car$, $\Omega \car\Theta$, $\Omega \car\Theta'$ and $\Omega \car\Theta\Theta'$. The corresponding orientifold planes can be either of type O6$^{(-,-)}$ with negative RR charge and tension or of the more exotic type O6$^{(+,+)}$ with positive RR charge and tension. For each of these four O6-planes, let us characterize the charges of these two kinds of crosscap states as $\eta_{\Omega R}$, $\eta_{\Omega R\Theta}$, $\eta_{\Omega R\Theta'}$ and $\eta_{\Omega R\Theta\Theta'}$, where $\eta_{\Om \car g}=+1$ refers to O6$^{(-,-)}$-planes and $\eta_{\Om \car g}=-1$ to O6$^{(+,+)}$-planes.

In the $\mbb{Z}_2 \times \mbb{Z}_2$ model without discrete torsion all four orientifold planes can be of type O6$^{(-,-)}$. In contrast, as pointed out in \cite{aaads99}, the perturbative orientifold with discrete torsion is forced to contain exotic O6$^{(+,+)}$-planes in order to satisfy the crosscap constraints. More concretely, the compatibility of the two relations
\bea
\langle \Omega \car |
e^{-l{\cal H}_{cl}} |
\Omega \car\, \Theta' \rangle &=& {\rm Tr}_{\theta'}  \left
(\Omega \car\, e^{-2\pi t H} \right), \nonumber \\
\langle \Omega \car\Theta |
e^{-l{\cal H}_{cl}} |
\Omega \car\, \Theta \Theta' \rangle &=& {\rm Tr}_{\theta'}  \left
(\Omega \car\, \Theta e^{-2\pi t H} \right),
\eea
as well as the choice of discrete torsion $\eta=\pm 1$ in the closed string sector enforce the relation
\be
\eta_{\Omega R}\,\eta_{\Omega R\Theta}\,\eta_{\Omega R\Theta'}\,\eta_{\Omega R\Theta\Theta'} = \eta
\label{opsigns}
\ee
among these various signs. Therefore, in the case with discrete torsion ($\eta = -1$), an odd number of O6$^{(+,+)}$-planes has to be present.
 
Working out the fixed point locus of the four orientifold projections $\Omega \car$, $\Omega \car\Theta$, $\Omega \car\Theta'$, $\Omega \car \Theta\Theta'$ and expressing everything in terms of bulk 3-cycles in the orbifold, we get
\bea 
\label{plane} 
\Pi_{O6} & = &
2\, \eta_{\Omega \car}\, [a^1]\cdot [a^2]\cdot [a^3] -
2^{1-2\beta^1-2\beta^2}\,  \eta_{\Omega \car\Theta}\, [b^1]\cdot [b^2]\cdot [a^3] \\
&& 
- 2^{1-2\beta^2-2\beta^3}\, \eta_{\Omega \car\Theta'}\, [a^1]\cdot [b^2]\cdot [b^3] -
2^{1-2\beta^1-2\beta^3}\, \eta_{\Omega \car\Theta\Theta'}\, [b^1]\cdot [a^2]\cdot [b^3] \nonumber
\eea
which is valid for any choice of discrete torsion $\eta$ and O6-plane charges.

Finally, we have to determine how $\Omega \car$ acts on the various 3-cycles. For the untwisted cycles this is straightforward: 
\be
\Omega \car:\cases{  [a^I]\to [a^I] \cr
                     [b^I]\to -[b^I] \cr}.
\ee
Therefore, the wrapping numbers are mapped as $\Omega \car:(n^I_a,\widetilde m^I_a)\to (n^I_a,-\widetilde m^I_a)$, respectively $\Omega \car:(n^I_a,m^I_a)\to (n^I_a,-2\beta^I \, n^I_a-m^I_a)$.

For the twisted sector 3-cycles, the canonical action corresponding to the models without vector structure would be $\Omega \car:\alpha^g_{ij,\,n}\to -\alpha^g_{\car(i)\car(j),\,n}$ and $\Omega \car:\alpha^g_{ij,\,m}\to \alpha^g_{\car(i)\car(j),\,m}$. However, we have to take into account that $\Omega \car$ acts in the $g$-twisted sector with an additional sign $\eta_{\Omega \car}\, \eta_{\Omega \car g}$ so that the final action reads 
\be
\Omega \car: 
\alpha^g_{ij,\,n}\to -\eta_{\Omega \car}\, \eta_{\Omega \car g} \,
\alpha^g_{\car(i)\car(j),\,n},\quad
\alpha^g_{ij,\,m}\to  \eta_{\Omega \car}\, \eta_{\Omega \car g}\,
\alpha^g_{\car(i)\car(j),\,m}  
\label{omegaR}
\ee
where for $\beta^I=0$ the reflection $\car$ leaves all fixed points $i\in\{1,2,3,4\}$ invariant, whereas for $\beta^I=1/2$ the action is given by
\be
\car:\cases{  1\to 1\cr
                                          2\to 2\cr
                                          3\to 4\cr
                                          4\to 3\cr }.
\ee
Note that these rules, including the extra minus signs in (\ref{omegaR})  is consistent with the conditions (\ref{bedi}). In fact, without them one would face an inconsistency.

\subsection{Spectrum}

The details of the orientifold action prove quite important in order to work out the massless spectrum of the low energy theory. In particular, they are essential in order to compute the chiral matter content which, while still described in terms of intersection numbers between 3-cycles, now includes chiral fermions transforming in the symmetric and anti-symmetric representations of $U(N)$ gauge groups. For simplicity, let us first consider D6-branes wrapping 3-cycles not invariant under $\car$, so that the gauge group is of the form $\prod_a U(N_a)$. In this case, we can apply a general rule for determining the massless left-handed chiral spectrum in terms of 3-cycles intersection numbers, as presented in table \ref{tcs}.
\begin{table}[htb]
\renewcommand{\arraystretch}{1.5}
\begin{center}
\begin{tabular}{|c|c|}
\hline
Representation  & Multiplicity \\
\hline
$\Yasymm_a$
%$[{\bf A_a}]_{L}$
 & ${1\over 2}\left(\Pi'_a\cdot \Pi_a
+\Pi_{{\rm O}6}\cdot \Pi_a\right)$  \\
$\Ysymm_a$
& ${1\over 2}\left(\Pi'_a\cdot \Pi_a-\Pi_{{\rm O}6} \cdot \Pi_a\right)$   \\
$(\antifund_a,\fund_b)$
& $\Pi_a\cdot \Pi_{b}$   \\
$(\fund_a,\fund_b)$
& $\Pi'_a\cdot \Pi_{b}$
\\
\hline
\end{tabular}
\caption{Chiral spectrum for intersecting D6-branes}
\label{tcs}
\end{center}
\end{table}

Given (\ref{prod}) and (\ref{inttwist}), it is now easy to compute the intersection number between two rigid D6-branes of the form (\ref{rigid}).  For instance, the intersection number of $\Pi_a^F$ with its $\Omega R$ image is given by
\bea
\left( \Pi_a^F\right)'\cdot \Pi_a^F & = & \eta_{\Omega R}\, 
\Bigl(
2 \eta_{\Omega R}\, \prod_I n^I_a\, \widetilde m_a^I -
   \eta_{\Omega R\Theta}\, {\textstyle {Q^1_{\beta^1}\, Q^2_{\beta^2}\over 2}}\, n^3\, \widetilde m_a^3 \\ \nonumber
&& \phantom{aaaa} 
- \eta_{\Omega R\Theta'}\, {\textstyle {Q^2_{\beta^2}\, Q^3_{\beta^3}\over 2}}\, n^1\, \widetilde m_a^1
-\eta_{\Omega R\Theta\Theta'}\, {\textstyle {Q^1_{\beta^1}\, Q^3_{\beta^3}\over 2}}\, n^2\, \widetilde m_a^2 \Bigr)
\eea
where $Q^I_{\beta^I}$ denotes the number of fixed points on the $I^{\rm th}$ $\T^2$ left invariant under the $\Omega \car$ action. This number is equal to two, except for the case that $\beta^I=1/2$ and $n_a^I=$ odd, where it is equal to one. For the intersection with the orientifold plane one obtains
\bea
\Pi_{O6}\cdot \Pi_a^F&=&2\, \eta_{\Omega R}\, \prod_I \widetilde m_a^I -
2^{1-2\beta^1-2\beta^2}\, \eta_{\Omega R\Theta}\,  n^1\, n^2\, \widetilde m^3
\\ \nonumber && 
- 2^{1-2\beta^2-2\beta^3}\, \eta_{\Omega R\Theta'}\, \widetilde m^1 \, n^2\, n^3
- 2^{1-2\beta^1-2\beta^3}\, \eta_{\Omega R\Theta\Theta'}\, n^1\, \widetilde m^2 \, n^3
\eea
and from both expressions one can obtain the chiral content involving symmetric and anti-symmetric representations of $\prod_a U(N_a)$.

Finally let us recall that, as stated above, those 3-cycles which are invariant under the orientifold action $\Om\car$ do not yield a unitary gauge group, but rather an orthogonal or symplectic group. More precisely, in the present case a stack of $N_a$ fractional D6-branes such that $\Om\car \Pi_a^F = \Pi_a^F$ yields a $USp(2N_a)$ gauge group. In the $\mbb{Z}_2 \ti \mbb{Z}_2'$ orientifold that we are discussing, the fractional D6-branes invariant under $\Om\car$ are those placed on top of an exotic O6$^{(+,+)}$-plane. For instance, if we choose $\eta_{\Om\car} = -1$ and $\eta_{\Om\car\Th} = \eta_{\Om\car\Th'} = \eta_{\Om\car\Th\Th'} = 1$ as crosscap charges, then we have 4 $\ti$ 64 different kinds of fractional D6-branes invariant under $\Om\car$, namely those which have bulk wrapping numbers $(1,0)(1,0)(1,0)$ and arbitrary twisted charges $\eps^g_{a,ij}$. On the other hand, a fractional D6-brane with bulk wrapping numbers $(1,0)(0,1)(0,-1)$ will still yield a $U(N_a)$ gauge group, since it is mapped to a fractional D-brane with same bulk wrapping numbers but different twisted charges. We will further illustrate the appearance of symplectic vs. unitary gauge groups with explicit examples in Section \ref{examples}. 

One should keep in mind that in addition to the topologically determined chiral matter there can appear extra non-chiral matter. The determination of the non-chiral spectrum needs a more thorough, case by case  analysis of the overlaps
between all possible pairs of  boundary states. For a concrete example, we will
come back to this point in section 5. 

\subsection{Consistency and supersymmetry conditions}

Having defined all the ingredients to construct intersecting D6-brane models on the $\mbb{Z}_2 \times \mbb{Z}_2'$ background, it only remains to recall that the consistency of the string construction imposes some constraints on the D6-brane configuration. In particular, we need to satisfy the cancellation of RR tadpoles. Part of RR tadpole cancellation can be easily expressed in terms of homology classes, namely
\be
\label{tadpole} 
\sum_a  N_a\, \left([\Pi_a] + [\Pi'_a]\right) = 4\,  \Pi_{O6}, 
\ee 
where $N_a$ denotes the number of branes wrapping the 3-cycle $\Pi_a$, as well as its $\Omega \car$ image  $\Pi'_a$.

However, it is important to recall that D-brane charges are often not fully classified by homology but rather by K-theory \cite{EW98}. Hence, as pointed out in \cite{AU00}, in order to construct a consistent model where RR tadpoles cancel we may also need to satisfy some extra constraints invisible to homology, which usually manifest themselves as anomalies on the world-volume of D-brane probes. It is easy to see that these extra K-theory constraints do appear in our $\mbb{Z}_2 \times \mbb{Z}_2'$ orientifold construction \cite{ms04a}, in the sense that D6-branes may carry torsion RR charges which are $\mbb{Z}_2$-valued, and which also need to vanish. In the present work, we will not attempt to derive these extra K-theory constraints in full generality and from first principles, leaving this task for future work. In the Appendix, however, we deduce them for a wide class of $\mbb{Z}_2 \ti \mbb{Z}_2'$ models, by using the D-brane probe arguments of \cite{AU00}.

Finally, we are intersted in constructing $\cn=1$ supersymmetric rigid D6-brane models. By the general discussion of Section \ref{moduli}, we know that we must require all the D6-branes to be calibrated by the holomorphic 3-form $\Om_3$ and with the same phase
\be
{\rm Vol\, } \left(\Pi_a\right) \, = \, \int_{\Pi_a} \preal \left(e^{i\phi_{O6}}\, \Om_3 \right), \quad \quad \forall a
\label{susy1}
\ee
where $\phi_{O6}$ is the calibration phase of the O6-plane, in our case $\phi_{O6} = 0$. In general ${\bf CY}_3$ compactifications, condition (\ref{susy1}) is equivalent to demanding that all the D6-branes are related by $SU(3)$ `rotations' with respect to the O6-plane. In the particular class of models at hand, which basically consist of D6-branes at three different angles $\th^1$, $\th^2$, $\th^3$ with the real axis of the three $\T^2$, this amounts to imposing the well-known condition \cite{bdl96}
\be
\th_a^1\, +\, \th_a^2\, +\, \th_a^3\, = \, 0\ {\rm mod}\ 2\pi, \quad \quad \forall a.
\label{susy2}
\ee

It turns out, however, that eq.(\ref{susy1}) is sometimes more useful in order to express the supersymmetry conditions in terms of closed string moduli. Indeed, notice that for $\phi_{O6} = 0$ this condition is equivalent to
\be
\int_{\Pi_a} \pim \left(\Om_3 \right) = 0, \quad \int_{\Pi_a} \preal \left(\Om_3 \right) > 0,
\label{susy3}
\ee
and that for factorizable D6-branes $\int \preal (\Om_3) = \preal \left( \int \Om_3 \right)$ and
\be
\int_{\Pi_a} \Om_3 = \frac{1}{4} \prod_{I=1}^3 \left(n_a^I R_1^I + i \widetilde{m}_a^I R_2^I \right), 
\label{cali}
\ee
where the $1/4$ factor arises from considering a fractional brane, and $R_1^I$, $R_2^I$ are as defined in fig. \ref{fct2}. Hence, eq.(\ref{susy1}) can be rewritten as
\be
\begin{array}{lcr}\vspace{.2cm}
\widetilde{m}_a^1\,  \widetilde{m}_a^2\,  \widetilde{m}_a^3 -
\sum_{I\ne J\ne K}  n_a^I\,  {n}_a^J\,  \widetilde{m}_a^K\, \left(U^I\, U^J\right)^{-1}&=&0 \\
n_a^1\,  n_a^2\,  n_a^3 -
\sum_{I\ne J\ne K}  n_a^I\,  \widetilde{m}_a^J\,  \widetilde{m}_a^K\, U^J\, U^K &>&0,
\end{array}
\label{susy4}
\ee
where we have introduced the three complex structure moduli $U^I=R^I_2/R^I_1$ of the three $\T^2$ factors.

\section{Examples}\label{examples}

Having described the set of rigid D6-branes which arise on the $\mbb{Z}_2 \times \mbb{Z}_2'$ orientifold, the associated low-energy spectrum and the conditions that they must satisfy, we are in position to build actual D-brane models which realize the scenario described in the introduction. The purpose of this section is to build such models explicitly, as well as discuss some of their properties. The aim is not to perform an exhaustive search for semi-realistic models in this context, which is beyond the scope of the present paper, but rather to give some simple examples that illustrate how chiral $\cn=1$ models via rigid intersecting D6-branes can be constructed. 

\subsection{Model Building strategy}\label{strategy}

So far, the discussion that we have made on rigid intersecting D-brane models on $\mbb{Z}_2 \times \mbb{Z}_2'$ is totally general. In order to construct some explicit $\cn=1$ chiral D-brane models, however, we will restrict to a subclass of the orientifold backgrounds discussed in the previous section, so that we can illustrate the model building possibilities of the present constructions in a simpler way.

Let us first consider a particular choice of orientifold plane content. As shown in Section \ref{intersecting}, the choice of discrete torsion on a $\mbb{Z}_2 \times \mbb{Z}_2$ orientifold implies some constraints on the O-planes charges, which is encoded in the relation (\ref{opsigns}). In the particular case at hand ($\eta =-1$), it implies that there is an odd number of O6-planes with positive tension. For definiteness, let us choose the crosscap charges to be $\eta_{\Om\car} = -1$ and $\eta_{\Om\car\Th} = \eta_{\Om\car\Th'} = \eta_{\Om\car\Th\Th'} = 1$. By a mirror symmetry transformation carrying intersecting to magnetized D-branes, this background translates into a $\mbb{Z}_2 \times \mbb{Z}_2$ type IIB orientifold without discrete torsion, and with $O3^{(+,+)}$ and $O7_i^{(-,-)}$-planes. In Section \ref{flux}, this fact will help us making contact with the flux vacua constructed in \cite{ms04a}.

In addition, we will further simplify the problem by restricting to orientifold backgrounds made only of untilted tori. That is, we will consider $(\T^2)^3/(\mbb{Z}_2 \times \mbb{Z}_2')$ backgrounds where each of the $\T^2$ factors has a rectangular complex structure, which is nothing but the choice $\b^I = 0, \ I=1,2,3$ in the language of figure \ref{fct2}.

Given these choices, and introducing fractional D6-branes of the form (\ref{rigid}), the RR tadpole conditions deduced from (\ref{tadpole}) can be expressed as a set of untwisted tadpole conditions given by
\be
\begin{array}{lll} \vspace*{.2cm}
\sum_a N_a n_a^1 n_a^2 n_a^3 & = & -16, \\\vspace*{.2cm}
\sum_a N_a m_a^1 m_a^2 n_a^3 & = & -16, \\\vspace*{.2cm}
\sum_a N_a m_a^1 n_a^2 m_a^3 & = & -16, \\\vspace*{.2cm}
\sum_a N_a n_a^1 m_a^2 m_a^3 & = & -16
\end{array}
\label{RRtad1}
\ee
plus the twisted ones, which take the form
\be
\begin{array}{lll} \vspace*{.2cm}
\sum_a N_a n_a^1 \eps_{a,ij}^{\Th'} & = & 0, 
\\ \vspace*{.2cm}
\sum_a N_a n_a^2 \eps_{a,jk}^{\Th\Th'} & = & 0,  
\\ \vspace*{.2cm}
\sum_a N_a n_a^3 \eps_{a,ki}^{\Th} & = & 0. 
\end{array}
\label{RRtad2}
\ee
Here we have $\eps_{a,ij}^{\Th'} \neq 0$ if and only if $ij \in  S_{\Th'}^a$, i.e., if the brane $a$ does goes through the fixed point $ij$ in the $\Th'$ twisted sector, etc. 

Actually, the conditions (\ref{tadpole}) are not enough to guarantee that a given model is consistent string theory construction. Indeed, in orientifold models cancellation of RR tadpoles goes beyond homology constraints of the form (\ref{tadpole}), and also implies the cancellation of K-theory torsion charges \cite{AU00}. These extra constraints have proven to be non-trivial in intersecting/magnetized D-brane models \cite{FM03,ms04a}, rendering some of the known models in the literature inconsistent. This is also the case for the $\mbb{Z}_2 \times \mbb{Z}_2'$ background, where such extra K-theory conditions turn out to be quite constraining. For the sake of clarity, we have deferred the discussion of such constraints to the Appendix. From the results in there, one concludes that any fractional D6-brane (\ref{rigid}) which does not have bulk wrapping numbers of the form
\be
(\rm{odd}, \rm{even}) \otimes (\rm{odd}, \rm{even}) \otimes (\rm{odd}, \rm{even})
\label{noK}
\ee
does have a non-trivial $\mbb{Z}_2$ torsion charge. Hence, in order to avoid uncanceled RR torsion charges, we will only consider D6-brane models where fractional D6-branes come in multiples of two. That is, where the number of D6-branes on each stack $a$ satisfies 
\be
N_a \in 2\mbb{N},\ \forall a
\label{noK2}
\ee
with the exception of D6-branes of the form (\ref{noK}). When faced to build semi-realistic models, this constraint naturally leads to consider vacua whose low energy spectrum contains a Pati-Salam gauge group and matter content. In the next subsections we will show explicit examples providing such spectra.

Now, even after taking these specific choices of orientifold background/D-brane content, it shows quite involved to solve conditions (\ref{RRtad1}), (\ref{RRtad2}) and (\ref{noK2}) at the same time. Compared to the model building difficulties of intersecting D6-branes on the $\mbb{Z}_2 \times \mbb{Z}_2$ orientifold with the opposite discrete torsion, the main new ingredient here comes from the RR twisted tadpoles (\ref{RRtad2}) that have now to be imposed. Roughly speaking, since the $\mbb{Z}_2 \times \mbb{Z}_2'$ orbifold possesses plenty of collapsed 3-cycles that D6-branes can wrap, the variety of RR charges (and thus the RR tadpoles constraints) greatly increases as compared to the $\mbb{Z}_2 \times \mbb{Z}_2$ constructions in \cite{csu01a,csu01}. Hence, it will prove useful to take some extra working assumptions, which will greatly simplify the twisted tadpole conditions but nevertheless allow us to construct a rich variety of models. 

The models that we will consider can be described as follows. Let us divide the total amount of D6-branes in a model in $K$ different sets. Each set $\a = a, b, c, \dots$ will consist of $J_\a$ stacks of D6-branes, $\a= \{ \a_1, \dots, \a_{J_\a} \}$. In turn, each stack $\a_\b$, $\b = 1, \dots, J_\a$, consist of $N_{\a_\b}$ coincident D6-branes, and gives rise to either a $U(N_{\a_\b})$ or a $USp(2N_{\a_\b})$ gauge group. Now, given a set $\a$  we will demand all the stacks $\a_\b$ belonging to it to be charged under the {\em same} fixed points of the $\mbb{Z}_2 \times \mbb{Z}_2'$ orbifold. More precisely, we will demand all the stacks $\a_\b \in \a$ to have the same $S^g_{\a_\b}$, $\forall \b$.
%%%%%%%%%%%%%%%%%%%%%%%%%%%%%%
\begin{figure}
\begin{center}
\epsfxsize=5.75in
\epsfbox{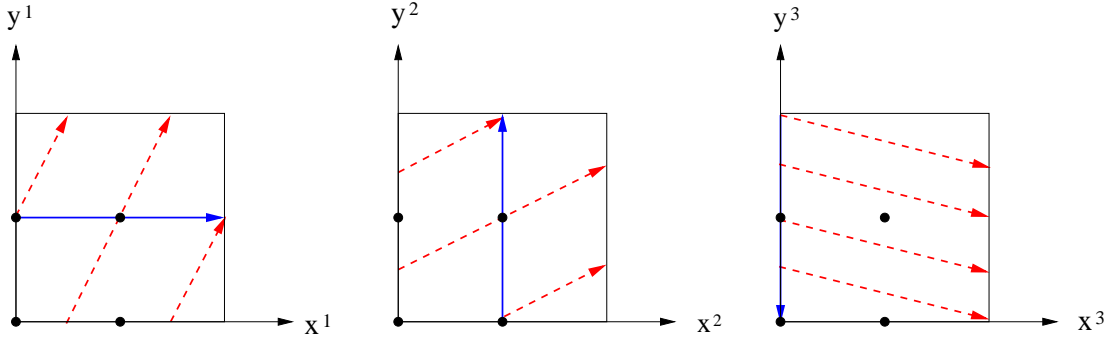}
\caption{Two fractional branes with different bulk wrapping numbers and charged under the same set of fixed points. \label{samefixed}}
\end{center}
\end{figure}
%%%%%%%%%%%%%%%%%%%%%%%%%%%%%%%
%

For instance, let us a consider a bunch of D6-branes with bulk wrapping numbers of the form 
\begin{center}
$(n_{\a_\b}^1, m_{\a_\b}^1) \otimes (n_{\a_\b}^2, m_{\a_\b}^2) \otimes (n_{\a_\b}^3, m_{\a_\b}^3) \, = \,
(\rm{odd}, \rm{even}) \otimes (\rm{even}, \rm{odd}) \otimes (\rm{even}, \rm{odd})$
\end{center}
for every choice of $\b = 1, \dots, J_\a$. Then, by the results of subsection 3.3 we know that if all these $J_\a$ stacks go through a particular fixed point of $\T^6/(\mbb{Z}_2 \times \mbb{Z}_2)$ (say the origin) they will share exactly the same set of fixed points (see figure \ref{samefixed}). Hence, each one will contribute to the same kind of twisted tadpoles, no matter which are the actual values of the $n$'s and the $m$'s describing each D-brane stack. 

If in addition we consider the same choice of discrete Wilson line for each stack $\a_\b$ of this set (say $\i =\i' = \i'' = 1$ in table \ref{signs}) the twisted tadpole conditions simplify to
\be
\begin{array}{lll} \vspace*{.2cm}
\sum_{\b = 1}^{J_\a} N_{\a_\b} n_{\a_\b}^1 & = & 0, \\\vspace*{.2cm}
\sum_{\b = 1}^{J_\a} N_{\a_\b} n_{\a_\b}^2 & = & 0,  \\\vspace*{.2cm}
\sum_{\b = 1}^{J_\a} N_{\a_\b} n_{\a_\b}^3 & = & 0.
\end{array}
\label{RRtad3}
\ee
and hence, by imposing (\ref{RRtad3}) the set $\a$ constructed above will have no net contribution to the twisted tadpoles. 

It is then clear that considering several sets $\a = a, b, c, d, \dots$ of the above form, the RR twisted tadpoles will be automatically canceled, and we only need to worry about the untwisted tadpoles (\ref{RRtad1}), as well as about the supersymmetry conditions (\ref{susy4}).

Given these facts, when building intersecting D-brane models in the next subsections, we will initially consider that every fractional D6-brane of a given model intersects the origin of $\T^6/(\mbb{Z}_2 \times \mbb{Z}_2)$, and that it has trivial discrete Wilson lines. This need not always be the case, and it will mainly serve to simplify the discussion below. We will comment on the additional freedom that we have in these class of models once some specific examples have been constructed.

\subsection{A simple $\cn=1$ D-brane model}

Recall that the $\mbb{Z}_2 \times \mbb{Z}_2'$ orientifold background being considered in the present paper was first discussed (in a T-dual version) in \cite{aaads99}, where it was an example of the class of {\em brane supersymmetry breaking} orientifolds. Given this, one may think that no $\cn=1$ D-brane vacua could ever be found for this background, in particular due to the presence of some O$p$$^{(+,+)}$-planes in the construction. However, such vacua do exist, as was shown in \cite{ms04a} by explicitly building simple $\cn=1$ examples via bulk D-branes. As a warm up before building more realistic models, let us review and generalize the simplest of such examples. Unlike in the rest of the section we will allow here for tilted two-tori with $\b^I \neq 0$.

In the language of intersecting D6-branes, the apparent obstruction to build $\cn=1$ vacua of \cite{aaads99} can be understood by the fact that RR and NSNS tadpoles can never be simultaneously satisfied if every D6-brane is (anti)parallel to an O6-plane. What was demonstrated in \cite{ms04a} is that solutions can be found when we allow for more general configurations, in particular those involving D6-branes at non-trivial angles with the O6-planes.

Indeed, let us consider a stack of $N$ (bulk) D6-branes with wrapping numbers
\be
\bigotimes_{I=1}^3\, (n^I, \widetilde{m}^I)\, = \, (-1,1-\beta^1) \otimes (-1,1-\beta^2) \otimes (-1,1-\beta^3),
\label{simplest}
\ee
which preserves $\cn=1$ supersymmetry if
\be
\arctan \left( (1-\beta^1)\, U^1 \right) + \arctan \left( (1-\beta^2)\, U^2 \right) + \arctan \left( (1-\beta^3)\, U^3 \right) = \pi.
\label{simplestN=1}
\ee

In this simple case, the twisted tadpole conditions are readily satisfied, since we are only dealing with bulk D-branes, and the untwisted ones can be solved for all values of $\beta^I$ by choosing $N=4$.\footnote{Recall that a bulk D-brane on $\mbb{Z}_2 \times \mbb{Z}_2'$ contributes to the untwisted tadpoles (\ref{RRtad1}) as 4 fractional D-branes.} The open string massless spectrum of this theory is given by a gauge group $U(4)$, as well as $32\beta^1 \beta^2 \beta^3$ chiral multiplets in the anti-symmetric representation. In addition, since we are wrapping our D6-branes on a bulk 3-cycle, there are also three chiral multiplets in the adjoint representation of $U(4)$.

Alternatively, we could introduce rigid branes with the same wrapping numbers, where it turns out that one needs four stacks and their $\Omega R$ images to cancel all twisted sector tadpoles. In this case we get the gauge group $U(4)^4$ and $8\beta^1 \beta^2 \beta^3$ chiral multiplets in the anti-symmetric representation of each $U(4)$ factor. However, this time we do not get any adjoint scalars.

Expanding the 3-cycles into a basis $\{\Sigma_i:i=1,\ldots,b_3\}$ as 
$\pi_a=\sum_i v_{a,i} \Sigma_i$, it is well known that via Green-Schwarz couplings of the form
\be
\label{green}
\sum_{a,i} \int N_a (v_{a,i}-v'_{a,i})\, F_a\wedge B_i
\ee
the initial $U(1)$ gauge factors can receive a mass term, surviving only as (perturbative) global symmetries. The massless $U(1)$'s are given by the kernel of the matrix $M_{a,i}=N_a (v_{a,i}-v'_{a,i})$. From these equations it is clear that for models with many 3-cycles it is much more unlikely for a $U(1)$ gauge boson to stay massless. 

In particular, in the above example one easily realizes that indeed all $U(1)$ gauge factors are massive, so that the surviving gauge symmetry is $SU(4)$, respectively $SU(4)^4$. Thus, strictly speaking, this type IIA models are not chiral vacua. More precisely, they are only chiral with respect to some global $U(1)$ symmetries. We now show, however, how $\cn=1$ chiral vacua can be constructed in this same setup.

\subsection{A chiral $\cn=1$ model via rigid intersecting branes}\label{split}

Let us now illustrate how a chiral $\cn=1$  compactification can be obtained, following the general model building strategy described in subsection \ref{strategy}. As before, we choose $- \eta_{\Om\car} = \eta_{\Om\car\Th} = \eta_{\Om\car\Th'} = \eta_{\Om\car\Th\Th'} = 1$ and $\beta^I=0$ for all $I$. It can thus be seen that the D6-brane content of table \ref{modelsplit} represents a type IIA $\cn=1$ chiral vacuum consisting of rigid intersecting D6-branes.

\begin{table}[htb]
\renewcommand{\arraystretch}{1.5}
\begin{center}
\begin{tabular}{|c||c|c|c|}
\hline
 $N_\a$  &  $(n_\a^{1},m_\a^{1})$  &  $(n_\a^{2},m_\a^{2})$
&  $(n_\a^{3},m_\a^{3})$ \\
\hline
\hline $N_{a_1} = 4$ & $(1,0)$ & $(0,-1)$ & $(0,1)$ \\
\hline $N_{a_2} = 4$ & $(-1,0)$ & $(0,1)$ & $(0,1)$ \\
\hline
\hline $N_{b_1}= 2$ & $(-1,1)$ &  $ (-2,1)$  & $(-2,1)$ \\
\hline $N_{b_2}= 2$ & $(1,-1)$ &  $ (2,-1)$  & $(-2,1)$ \\
\hline $N_{b_3}= 2$ & $(-1,1)$ &  $ (2,-1)$  & $(2,-1)$ \\
\hline $N_{b_4}= 2$ & $(1,-1)$ &  $ (-2,1)$  & $(2,-1)$ \\
\hline
\hline $N_{c_1}= 4$ & $(1,0)$ &  $ (1,0)$  & $(1,0)$ \\
\hline $N_{c_2}= 4$ & $(-1,0)$ &  $ (-1,0)$  & $(1,0)$ \\
\hline $N_{c_3}= 4$ & $(-1,0)$ &  $ (1,0)$  & $(-1,0)$ \\
\hline $N_{c_4}= 4$ & $(1,0)$ &  $ (-1,0)$  & $(-1,0)$ \\
\hline
\end{tabular}
\caption{\small Wrapping numbers of a rigid, $\cn=1$, chiral model containing a Pati-Salam four-family spectrum. \label{modelsplit}}
\end{center}
\end{table}

Let us describe the D6-brane model of table \ref{modelsplit} in some detail. The model consist of 3 sets of D6-branes, indexed by $\a = a,b,c$, and each of these sets consist of several stacks of fractional D6-branes $a_1$, $a_2$, etc. Notice that we are using a compact notation to describe fractional D6-branes of the form (\ref{rigid}), by only specifying the bulk wrapping numbers $(n_\a^i,m_\a^i)$ of each fractional D-brane.\footnote{With the understanding that each of the wrapping numbers in (\ref{ineqbulk}) represent a different fractional D-brane. See below.} This notation is well-defined because, as explained before, we have chosen each fractional D6-brane to intersect the origin of $\T^2 \times \T^2 \times \T^2$ and to have trivial discrete Wilson lines. These conventions understood, it is easy to derive the signs $\eps_{a,ij}^g$ in (\ref{rigid}). For instance, the fractional branes of the set $a$ correspond to
\be
\begin{array}{rcl}
\Pi^B_{a_1} = (1,0)(0,-1)(0,1) 
& \raw & 
\Pi^F_{a_1} = {1 \over 4} \Pi^B_{a_1} + {1 \over 4} 
\left( \sum_{i,j\in (13)\times (12)} \alpha^\Theta_{ij,\,m} \right) \\ \vspace*{.25cm}
& &
+ {1\over 4} 
\left(\sum_{j,k\in (12)\times (12)} \alpha^{\Theta'}_{jk,\,n} \right) 
- {1\over 4} 
\left( \sum_{i,k\in (13)\times (12)} \alpha^{\Theta\Theta'}_{ik,m} \right)
\\
\Pi^B_{a_2} = (-1,0)(0,1)(0,1) 
& \raw & 
\Pi^F_{a_2} = {1 \over 4} \Pi^B_{a_2} + {1 \over 4} 
\left(\sum_{i,j\in (13)\times (12)} \alpha^\Theta_{ij,\,m} \right) \\
& &
- {1\over 4} 
\left(\sum_{j,k\in (12)\times (12)} \alpha^{\Theta'}_{jk,\,n} \right) 
+ {1\over 4} 
\left( \sum_{i,k\in (13)\times (12)} \alpha^{\Theta\Theta'}_{ik,m} \right)
\end{array}
\label{dict}
\ee
and similarly for the other fractional D-branes in table \ref{modelsplit}. 

To the D6-brane content of table \ref{modelsplit} we need to add the images under the orientifold action $\Om\car$. In the case of the set $a$ these correspond to
\be
\begin{array}{rcl}
\Pi^B_{a_1} = (1,0)(0,-1)(0,1) 
& \stackrel{\Om\car}{\longraw} & 
\Pi^B_{a_1^\prime} = (1,0)(0,1)(0,-1),
\\
\Pi^B_{a_2} = (-1,0)(0,1)(0,1) 
& \stackrel{\Om\car}{\longraw} & 
\Pi^B_{a_2^\prime} = (-1,0)(0,-1)(0,-1).
\end{array}
\label{images}
\ee
Notice that, in this particular case, all the fractional D6-branes $a_i, a_i'$ correspond to the same bulk homology 3-cycle, i.e, $\Pi_{a}^B = \Pi_{a_1}^B = \Pi_{a_2}^B = \Pi_{a_1'}^B = \Pi_{a_2'}^B$. They differ, however, in the twisted homology 3-cycles that they wrap, as can be appreciated from (\ref{dict}). Notice, as well, that when all the RR/homology charges of this set are added the total twisted charge vanishes, leaving only a net bulk D6-brane charge proportional to $\Pi_{a}^B$. 

An analogous effect occurs for the other two sets $b$ and $c$. Indeed, again the twisted RR charges cancel among the components of each set (plus their orientifold images), leaving a net bulk D-brane charge only. This clearly matches with the fact that the conditions (\ref{RRtad3}) are satisfied for the sets $a$, $b$ and $c$ separately. On the other hand, adding up the untwisted RR charges of these three sets one can check that the conditions (\ref{RRtad1}) are also satisfied, so that we are considering a consistent type IIA D-brane model free of RR tadpoles. Finally, the $\cn=1$ supersymmetry conditions are also satisfied if we choose
\be
\arctan \left(U^1\right) + \arctan \left(U^2/2 \right) + \arctan \left(U^3/2\right) = \pi.
\label{susysplit}
\ee
so that we recover an $\cn=1$ supersymmetric type IIA vacuum.

A closer look to the D6-brane configuration of table \ref{modelsplit} reveals that, from the model building point of view, it can be understood in terms of three stacks of pure bulk D6-branes at angles, wrapping $\Pi_a^B$, $\Pi_b^B$ and $\Pi_c^B$, and which cancel the untwisted RR tadpoles of the orientifold background. Such bulk D-branes have been chosen to go through the orbifold singularities, and hence split into the four fractional constituents of the regular representation of $\mbb{Z}_2 \times \mbb{Z}_2$. In this sense, the model of table \ref{modelsplit} is analogous to the constructions of D-branes at orbifold singularities of \cite{bl96,abpss96a,ks97a,ks97b,afiv98}.

Let us now compute the spectrum of this model. Since the D6-branes ${a_i}$, $i=1,2$ differ from their orientifold images ${a_i'}$ in their twisted cycles, each of them yields a unitary gauge group $U(N_{a_i})$ at low energies. The same applies to each of the fractional D-branes of the set $b$. On the other hand, the fractional D6-branes $c_i$ are fixed by the orientifold action and so, when performing the orientifold projection on the Chan-Paton degrees of freedom, we recover a $USp(2N_{c_i})$ gauge group. The resulting gauge group is naively
\be
U(4)^2 \times U(2)^4 \times USp(8)^4.
\label{gaugesplit}
\ee
However, taking the Green-Schwarz couplings (\ref{green}) into account one finds that all of the $U(1)$ gauge bosons acquire a mass, so that at the end of the day the gauge group is $SU(4)^2 \times SU(2)^4 \times USp(8)^4$. 

The chiral spectrum can be obtained by means of computing the intersection numbers between each pair of D6-branes, and using table \ref{tcs}. The result is presented in table \ref{chiralsplit}. One can check that, as expected from RR tadpole cancellation, the non-abelian gauge anomalies cancel.
\begin{table}[htb]
\renewcommand{\arraystretch}{1.15}
\begin{center}
\begin{tabular}{|c|c|}
\hline
Sector  &   Rep. of $U(4)^2\times U(2)^4 \times USp(8)^4$ \\
\hline
$(a_1b_1)$  &  $4\times (\o{4},1;2,1,1,1;1,1,1,1)$  \\
$(a_1'b_2)$  & $4\times ({4},1;1,2,1,1;1,1,1,1)$  \\
$(a_2b_3)$  & $4\times (1,\o{4};1,1,2,1;1,1,1,1)$  \\
$(a_2'b_4)$    & $4\times (1,{4};1,1,1,2;1,1,1,1)$  \\
\hline
$(b_1c)$ &  $1\times (1,1;\o{2},1,1,1;8,1,1,1)$  \\
$(b_2c)$ &  $1\times (1,1;1,\o{2},1,1;1,8,1,1)$  \\
$(b_3c)$ & $1\times (1,1;1,1,\o{2},1;1,1,8,1)$  \\
$(b_4c)$ & $1\times (1,1;1,1,1,\o{2};1,1,1,8)$  \\
\hline
$(b_1'b_2)$ &  $2\times (1,1;2,2,1,1;1,1,1,1)$  \\
$(b_3'b_4)$ & $2\times (1,1;1,1,2,2;1,1,1,1)$  \\
$(b_1'b_3)$ & $6\times(1,1;2,1,2,1;1,1,1,1)$  \\
$(b_1'b_4)$ &  $6\times(1,1;2,1,1,2;1,1,1,1)$  \\
$(b_2'b_3)$ & $6\times(1,1;1,2,2,1;1,1,1,1)$  \\
$(b_2'b_4)$ & $6\times(1,1;1,2,1,2;1,1,1,1)$  \\
\hline
$(b_1'b_1)$ &  $18\times (1,1;\Yasymm,1,1,1;1,1,1,1)$  \\
$(b_2'b_2)$ & $18\times (1,1;1,\Yasymm,1,1;1,1,1,1)$  \\
$(b_3'b_3)$ & $18\times (1,1;1,1,\Yasymm,1;1,1,1,1)$  \\
$(b_4'b_4)$ & $18\times (1,1;1,1,1,\Yasymm;1,1,1,1)$  \\
\hline
\end{tabular}
\caption{Chiral spectrum of the D6-brane model of table \ref{modelsplit}. \label{chiralsplit}}
\end{center}
\end{table}

To be accurate, in table \ref{chiralsplit} we should have included the spectrum arising from the $ac$ sectors, which we now display in table \ref{vectorsplit}.
\begin{table}[htb]
\renewcommand{\arraystretch}{1.15}
\begin{center}
\begin{tabular}{|c|c|}
\hline
Sector  &   Rep. of $U(4)^2\times U(2)^4 \times USp(8)^4$ \\
\hline
$(a_1c_2)$  & $(4,1;1,1,1,1;1,8,1,1)$  \\
$(a_1c_3)$  & $(\o{4},1;1,1,1,1;1,1,8,1)$  \\
$(a_2c_1)$  & $(1,\o{4};1,1,1,1;8,1,1,1)$  \\
$(a_2c_4)$  & $(1,4;1,1,1,1;1,1,1,8)$  \\
\hline
\end{tabular}
\caption{Extra chiral spectrum of the model of table \ref{modelsplit}. \label{vectorsplit}}
\end{center}
\end{table}
The reason that we have not included this part of the spectrum is that we can consider a simple deformation of our model which eliminates this massless sector of the theory. Indeed, instead of considering that the D-branes $c_i$ go through the origin of $\T^2 \times \T^2 \times \T^2$, we can place them at another fixed point location. For instance, we could think that the rigid D-branes $c_i$, $i=1,2,3,4$, intersect the fixed points 2 and 4 in in the first $\T^2$, so that the intersection numbers with any D-brane of the set $a$ do now vanish. Alternatively, the same effect can be obtained by turning on a non-trivial discrete Wilson line on the first $\T^2$, again for each of the D6-branes $c_i$. As a result, the massless spectrum of table \ref{vectorsplit} no longer arises after this discrete deformation.

In addition to this chiral spectrum there is, of course, light non-chiral matter in this model. By construction all the 3-cycles that the D6's wrap are rigid, and so for these branes there are no scalars transforming in the adjoint representation. However, just as it happens in models of branes at $\mbb{Z}_2 \times \mbb{Z}_2$ singularities, additional bifundamental non-chiral fields appear between two fractional D-branes. We will further comment on this point and explain its consequences in Section \ref{gaugino}.

\subsection{An $\cn=1$ Pati-Salam-like example}
\label{nonsplit}

The purpose of the previous models was to exemplify how $\cn=1$ chiral compactifications can be achieved by means of rigid intersecting D6-branes, in the particular framework of a $\mbb{Z}_2 \times \mbb{Z}_2'$ orientifold of type IIA string theory. These vacua are, nevertheless, too simple, since each of the sets $\a = a, b, c\dots$ of rigid D6-branes can be understood as a bulk (non-rigid) D6-brane $\a$ placed on a fixed point set of the orbifold action, and split into fractional D6-branes $\a_1, \a_2, \a_3, \a_4$ in the regular representation of $\mbb{Z}_2 \times \mbb{Z}_2'$. As a result, this class of constructions suffer from a serious drawback when trying to build semi-realistic D-brane models. First, as can be appreciated from (\ref{gaugesplit}) the low energy gauge group will have only factors of the form $U(N)^2$, $U(N)^4$ and $USp(2N)^4$, from which it seems difficult to build a semi-realistic spectrum. Second, although we have got rid of the adjoint fields by splitting the bulk D6-branes into its fractional constituents, there will still be extra, non-chiral degrees of freedom which smoothly connect this configuration with one consisting of non-rigid D6-branes (see Section \ref{gaugino}), and where the adjoint matter fields reappear. 

Nevertheless, one of the lessons that we learn from building models of D-branes at singularities is that there exist a wide class of possibilities for building $\cn=1$ chiral models, of which arranging fractional D-branes in regular representations is just the most obvious one \cite{aiqu00}. This is also the case for fractional intersecting D-branes which, after some T-dualities, can be seen as a generalization of models of D-branes at orbifold singularities in either type I or type IIB orientifolds. Indeed, the model building strategy described at the beginning of this section applies to a much richer class of rigid D-brane models other than fractional D-branes in the regular representation. The purpose of the present subsection is to give an example of these more involved constructions which, as we presently show, also admit low energy theories much closer to realistic particle physics.

\begin{table}[htb]
\renewcommand{\arraystretch}{1.5}
\begin{center}
\begin{tabular}{|c||c|c|c|}
\hline
 $N_\a$  &  $(n_\a^{1},m_\a^{1})$  &  $(n_\a^{2},m_\a^{2})$
&  $(n_\a^{3},m_\a^{3})$ \\
\hline
\hline $N_{a_1} = 4$ & $(1,0)$ & $(0,1)$ & $(0,-1)$  \\
\hline $N_{a_2} = 2$ & $(1,0)$ & $(2,1)$ & $(4,-1)$  \\
\hline $N_{a_3} = 2$ & $(-3,2)$ & $(-2,1)$ & $(-4,1)$  \\
\hline
\hline $N_{b_1}= 2$ & $(1,0)$ &  $ (0,1)$  & $(0,-1)$ \\
\hline $N_{b_2}= 2$ & $(-1,0)$ &  $ (0,1)$  & $(0,1)$ \\
\hline
\hline $N_{c_1}= 4$ & $(0,1)$ &  $ (1,0)$  & $(0,-1)$ \\
\hline $N_{c_2}= 4$ & $(0,1)$ &  $ (-1,0)$  & $(0,1)$ \\
\hline
\hline $N_{d_1}= 2 N_f$ & $(1,0)$ & $(1,0)$ & $(1,0)$  \\
\hline $N_{d_2}= 2 N_f$ & $(1,0)$ & $(-1,0)$ & $(-1,0)$  \\
\hline $N_{d_3}= 2 N_f$ & $(-1,0)$ & $(1,0)$ & $(-1,0)$  \\
\hline $N_{d_4}= 2 N_f$ & $(-1,0)$ & $(-1,0)$ & $(1,0)$  \\
\hline
\end{tabular}
\caption{\small Wrapping numbers of a rigid, $\cn=1$, chiral model containing a Pati-Salam four-family spectrum.\label{modelnonsplit}}
\end{center}
\end{table}

We present such example in table \ref{modelnonsplit}, which now consist of four sets of D6-branes, $\a = a, b, c, d$. The semi-realistic gauge group and chiral content of the theory will arise from the set $a$, which consist of three stacks of fractional D6-branes, each with {\em different} bulk wrapping numbers. Hence, unlike in the previous examples, this set of fractional branes cannot be smoothly deformed to a bulk D-brane. On the other hand, the sets $b$, $c$ and $d$ can still be seen as a bulk D6-brane split into fractional components. It is easy to check that this configuration satisfies the RR twisted tadpole constrains (\ref{RRtad3}), as well as the untwisted constraints (\ref{RRtad1}) upon choosing $N_{f} = 2$. Finally, the supersymmetry conditions (\ref{susy2}) amount, in this case, to
\be
\begin{array}{c}\vspace*{.2cm}
2 U^2 = U^3, \\
\arctan \left(2U^1/3\right) + \arctan \left(U^2/2\right) + \arctan \left(U^3/4\right) = \pi,
\end{array}
\label{susynonsplit}
\ee
which indeed have many solutions.

The gauge group derived from the D-brane set $a$ is given by $U(4) \times U(2) \times U(2)$ and, as we will now see, it can be seen as a sector yielding a Pati-Salam-like theory. On the other hand, the sets $b$, $c$ and $d$ yield, respectively, the gauge groups $U(2)^2$, $U(4)^2$ and $USp(4N_f)^4$. Notice that, by taking some flat directions, we can deform these latter fractional D-branes into bulk D-branes. The change of the gauge group upon this Higgsing is given by
\bea
\{b_1, b_2\},\ \{c_1, c_2\}, \ \{d_1, d_2, d_3, d_4\} & \raw & b,\ c,\ d \\  
%\nonumber
U(2)^2 \times U(4)^2 \times USp(2N_f)^4 & \raw & U(1) \times U(2) \times USp(2N_f).
\label{higgsing}
\eea
Again, some of the $U(1)$ factors of this group are not really gauge symmetries, but only global ones, since their would-be gauge boson receives a Stueckelberg mass by means of the Green-Schwarz coupling (\ref{green}). Taking this into account we recover a Pati-Salam gauge group $SU(4) \times SU(2) \times SU(2)$ from the set $a$, and an extra gauge group $SU(2)^2 \times SU(4)^2 \times USp(2N_f)^4$ from the other three sets when split into fractional D-branes. After the Higgsing (\ref{higgsing}), nevertheless, we still get a gauge group $U(1) \times U(2) \times USp(2N_f)$.

\begin{table}[htb!]
\renewcommand{\arraystretch}{1.3}
\begin{center}
\begin{tabular}{|c|c|c|}
\hline Sector & $U(4) \ti U(2) \ti U(2)$ & $U(2)^2 \ti U(4)^2 \ti USp(4N_f)^4$ \\
\hline
\hline ($a_1a_2$) &  $2(\overline 4,2,1)$ & $(1,1;1,1;1,1,1,1)$\\
\hline ($a_1'a_2$) &  $2(\overline 4,\overline 2,1)$ & $(1,1;1,1;1,1,1,1)$\\
\hline ($a_1a_3$) &  $4({4},1,\overline 2)$ & $(1,1;1,1;1,1,1,1)$\\
\hline ($a_2a_3$) &  $6(1,\overline 2,{2})$ & $(1,1;1,1;1,1,1,1)$\\
\hline ($a_2'a_3$) &  $10(1,{2},{2})$ & $(1,1;1,1;1,1,1,1)$\\
\hline ($a_2'a_2$) &  $4(1,\o\Yasymm,1)$ & $(1,1;1,1;1,1;1,1)$\\
\hline ($a_3'a_3$) &  $24(1,1,\o\Ysymm) + 96(1,1,\o\Yasymm)$ & $(1,1;1,1;1,1,1,1))$\\
\hline
\hline ($a_3b_1$) &  $4(1,1,2)$ & $(\overline 2,1;1,1;1,1,1,1)$\\
\hline ($a_3'b_2$) &  $12(1,1,\overline 2)$ & $(1,\overline 2;1,1;1,1,1,1)$\\
\hline ($a_2'c_1$) &  $(1,\overline 2,1)$ & $(1,1;\overline 4,1;1,1,1,1)$\\
\hline ($a_2c_2$) &  $3(1,\overline 2,1)$ & $(1,1;1,4;1,1,1,1)$\\
\hline ($a_3c_1$) &  $2(1,1,\overline 2)$ & $(1,1;4,1;1,1,1,1)$\\
\hline ($a_3'c_1$) &  $3(1,1,\overline 2)$ & $(1,1;\overline 4,1;1,1,1,1)$\\
\hline ($a_3c_2$) &  $(1,1,\overline 2)$ & $(1,1;1,4;1,1,1,1)$\\
\hline ($a_3'c_2$) &  $6(1,1,\overline 2)$ & $(1,1;1,\overline 4;1,1,1,1)$\\
\hline ($a_3d_1$) &  $2(1,1,{2})$ & $(1,1;{1,1;4N_f},1,1,1)$\\
\hline
\end{tabular}
\caption{\small Four generation Pati-Salam-like $\cn=1$ spectrum derived from the D-brane content of table \ref{modelnonsplit}. Here we consider $b$, $c$ and $d$ being split into fractional D-branes, and none of them going through the origin.\label{chiralnonsplit}}
\end{center}
\end{table}

Let us now consider the chiral spectrum of this theory, which is presented in table \ref{chiralnonsplit}. For completeness, we have kept all the massive $U(1)$ factors from the gauge group, since they are still perturbative global symmetries of the theory and hence play an important role in the low energy dynamics. 

The most interesting part of the spectrum arises form the set $a$ of D-branes, which not only yields a $SU(4) \times SU(2) \times SU(2)$ gauge group but also contains the chiral fermions transforming as $4(4,1,2) + 4(\overline 4, 2,1)$. This is nothing but a four-family Pati-Salam spectrum. Notice that these are the only chiral fermions charged under the non-Abelian gauge group $SU(4)$, and that they are singlets under the extra gauge group $U(2)^2 \ti U(4)^2 \ti USp(4N_f)^4$. The Higgs sector of the theory also arises from the set $a$ although, in this case, is far from minimal.

Although the sets $b$, $c$ and $d$ do not give any chiral spectrum by themselves, they do intersect non-trivially the set $a$ and give extra chiral matter charged under $SU(4) \ti SU(2) \ti SU(2)$. However, just as done previously, by appropriately placing the sets $b$, $c$ and $d$ at fixed points away from the origin, we can minimize the amount of extra chiral matter. This minimal chiral matter content corresponds to the spectrum shown in table \ref{chiralnonsplit}. In particular, notice that there is no extra chiral matter charged under $SU(4)$.

Finally, not only does this $\cn=1$ model have an appealing chiral spectrum but, as we will see in the next section, is specially interesting for its massless non-chiral spectrum. As can be advanced from the general discussion above, the gauge groups $SU(4) \ti SU(2) \ti SU(2)$ do not have any adjoint fields charged under them, as they arise from rigid D6-branes. It will turn out, in addition, that the extra non-chiral spectrum that could be charged under this gauge group is also quite minimal, yielding an asymptotically-free $SU(4)$ gauge theory.

\section{Asymptotic freedom}
\label{gaugino}

The main motivation for getting rid of adjoint multiplets was to obtain a low-energy asymptotically free gauge theory with negative one-loop beta-function. This both can improve the running of the gauge couplings towards the converging running in the MSSM, \footnote{The issue of gauge unification in intersecting D-brane models has been considered in \cite{bls03}.} as well as allows for a gaugino condensate to form, via the non-perturbative superpotential
\be
W_{eff}\sim {M_s\, \beta^g_1 \over 32 \pi^2} \exp\left( {8\pi^2\over g_{YM}^2\,  \beta^g_1} \right).
\ee
Here the gauge couplings depend on the complex (K\"ahler) structure moduli in Type IIA (IIB) string theory. This effective potential could, in principle, be used to freeze some of the closed string moduli of the theory,\footnote{For the $\cn=1$ intersecting D-brane models of \cite{csu01,csu01a}, such effects have been studied in \cite{clw03}.} in addition to some extra sources of moduli stabilization like, e.g., fluxes.

When considering a chiral $\cn=1$ model based on completely rigid D-branes, by construction no massless adjoint fields will appear in the spectrum. This will, in principle, improve the beta function behavior towards asymptotic freedom. However, one must be careful since beta functions are sensitive to the entire massless spectrum, including some extra light non-chiral states. In general, the latter cannot be computed via intersection numbers or other topological invariants, so it proves important to have a good control of the whole spectrum of the theory. Toroidal orientifolds provide us with a particularly treatable set of $\cn=1$ D-brane vacua, since we can make use of BCFT techniques in order to compute the massless, non-chiral open string spectrum of the theory. Let us exemplify the lesson one learns from computing the entire light spectrum with a few chiral models from the previous section.

\vskip 0.5cm
\noindent
{\bf Splitting bulk branes:}
\vskip 0.2cm

Let us first consider a bulk D-brane, $a$,  with gauge group $U(N)$. This bulk brane has three adjoint chiral multiplets of $U(N)$ and, in addition other charged matter which arises from, e.g., the intersection with other D-branes. Ignoring this additional matter, the contribution to the beta-function reads $b_1^{U(N)} =-(3N) + 3\times N=0$, so we find that the beta functions of bulk D-branes are either vanishing or positive. 

Now, one may split the bulk brane into four rigid constituents $\{b_1,b_2,b_3,b_4\}$ in the regular representation of $\mbb{Z}_2 \ti \mbb{Z}_2$ as, e.g., the D-brane set $b$ in table \ref{modelsplit}. This leads to a gauge group $U(N)^4$ without massless adjoints. However, additional non-chiral matter can in principle appear between pairs of the four fractional D-branes, and this in fact  turns out to be the case.

By standard arguments, computing the overlap of two such boundary states
\be 
\label{overl}
\tilde A_{b_i,b_j}=    \int_0^\infty  dl\, \langle b_i| e^{-2\pi l H_{cl}}  |b_j \rangle + \int_0^\infty  dl\, \langle b_j| e^{-2\pi l H_{cl}}  |b_i \rangle, \quad\quad  i\ne j
\ee
the different signs for the twisted sector part imply that in the loop
channel amplitude  
\be
A_{b_i,b_j}=\int_0^\infty  {dt\over t}\, {\rm Tr}_{ij+ji} \left({\textstyle {1 + \Theta + \Theta ' + \Theta\Theta' \over 4}} \, e^{-2\pi t H_{o}} \right) 
\ee
only one hypermultiplet appears at the massless level. The scalars in this hypermultiplet  are associated with one pair of the oscillators $\psi^I_{-{1\over 2}}|0\rangle$, $\o\psi^I_{-{1\over 2}}|0 \rangle$, $I\in \{1,2,3\}$, where the $\psi^I$ are the world-sheet superpartners of the bosonic fields $Z^I$.  Note that for the overlap between two identical fractional branes, i.e. $i=j$ in (\ref{overl}), the modes $\psi^\mu_{-{1\over 2}}|0 \rangle$, $\o\psi^\mu_{-{1\over 2}}|0 \rangle$ of the non-compact oscillators survive the projection, leading to an ${\cal N}=1$ vector multiplet. 

Therefore with respect to each $U(N)$ factor, there exist $6N$ chiral supermultiplets in the fundamental representation, so that the one-loop beta-function is still $b_1^{U(N)} = -(3N) + 6N \times {1\over 2}=0$. It is easy to see that there is no actual improvement for the beta function when splitting a $U(N)$ bulk D-brane into its fractional constituents. As a consequence, any rigid D-brane obtained in this way (as the ones in the model of table \ref{modelsplit}) could not possibly yield a gauge group which is asymptotically free.

\vskip 0.5cm
\noindent
{\bf Generic rigid branes:}
\vskip 0.2cm

Splitting bulk D-branes into $\mbb{Z}_2 \ti \mbb{Z}_2$ fractional constituents is the simplest way of obtaining a rigid D-brane model, but not the only or most general one. For instance, we can consider the D-brane $a_1$ in table \ref{modelnonsplit}, and which yields a $U(4)$ gauge group. Let us check whether the $SU(4)$ gauge symmetry derived from this rigid D-brane is asymptotically free or not.

The chiral spectrum of this model is listed in table \ref{chiralnonsplit}. The sector charged non-trivially under $SU(4)$ solely consists of four families of a Pati-Salam theory. Regarding the non-chiral matter, since the brane is rigid there are no additional states in the adjoint representation. Nevertheless, by means of analogous computations to the one performed above for splitted bulk D-branes, from the $a_1a_1'$ sector we find one hypermultiplet in the antisymmetric representation of $SU(4)$.

With this particle content the $SU(4)$ gauge theory is asymptotically free, since the corresponding beta function reads
\be
b_1^{SU(4)}=-3 \times 4 + 16 \ti \half  + 2 \ti 1 = -2
\label{bSU(4)}
\ee
which is still negative. In the following we will show that there are no other light states charged under $SU(4)$, and hence no further contributions to (\ref{bSU(4)}).

Indeed, the only additional massless states charged under $SU(4)$ may arise from non-chiral matter, and more precisely from bifundamental vector-like pairs in the $a_1\a_j$ sector, where $\a_j$ is an arbitrary fractional D-brane. As mentioned earlier, for computing the non-chiral massless spectrum one has to keep in mind that the twisted sector parts of boundary states lead to the various $\mbb{Z}_2$ projections in loop channel. As a result, one can compute the spectrum between two fractional D-branes by considering the whole spectrum between these two in the unorbifolded theory and then performing the orbifold projection. 

\begin{figure}
\begin{center}
\epsfbox{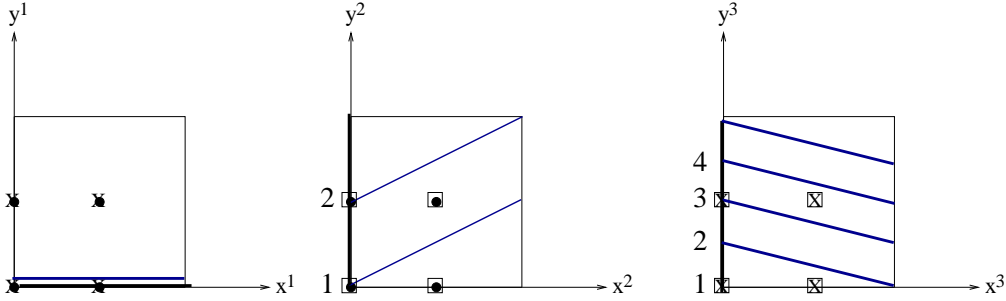}
\caption{Intersection points of the rigid branes $a_1$ and $a_2$. 
The numbers on the second and third $\T^2$ count the eight intersection points.}
\label{fct3}
\end{center}
\end{figure}

Let us illustrate this for the open string sector arising between the branes $a_1$ and $a_2$ from table \ref{modelnonsplit}. The bulk wrapping numbers of these D-branes are
\bea
 a_1=(1,0)(0,1)(0,-1), \quad\quad a_2=(1,0)(2,1)(4,-1),
\eea
so that they are lying on top of each other on the first $\T^2$, and therefore the chiral intersection number in the untwisted sector vanishes (yielding 
an ${\cal N}=2$ open string sector). The positions of these D-branes are displayed in figure \ref{fct3}.
%
%%%%%%%%%%%%%%%%%%%%%%%%%
%%%%%%%%%%%%%%%%%%%%%%%%%
%

The complete intersection number including also the twisted sectors reads
\bea
\label{inters}
I_{a_1,a_2}=\sum_g I^g_{a_1,a_2}={1\over 4}(0) +{1\over 4}(16) +{1\over 4}(0) +{1\over 4}(-8)=2,
\eea
where $g$ runs over all untwisted respectively twisted sectors $g\in\{1,\Theta,\Theta',\Theta\Theta'\}$. The question is whether there exist additional non-chiral states in this sector. Due to the non-trivial intersections on the second and third $\T^2$ factor, in the untwisted sector we get  eight hypermultiplets (chiral + anti-chiral). The $\Theta'$ sector in (\ref{inters}) is also vanishing, which indicates that  $\Theta'$ acts on complete  hypermultiplets, i.e., without projecting out chiral components. Indeed $\Theta'$ acts on the eight hypermultiplets  as
\bea
 \Theta': && (H_{11},H_{21},H_{13},H_{23},H_{12},H_{22},H_{14},H_{24})\to
 \nonumber \\
  && (-H_{11},-H_{21},-H_{13},-H_{23},-H_{14},-H_{24},-H_{12},-H_{22})
\eea
where the subscripts denote the eight intersection points of the two branes (see figure \ref{fct3}). The minus sign is due to the different signs of the second and fourth term in (\ref{inters}). 

Next we have to determine the action of $\Theta$ on the eight hypermultiplets. From (\ref{inters}) one deduces that this operation alone would lead to eight chiral multiplets. This is in accordance with the fact that all fixed points are invariant under $\Theta$ and that $\Theta$ kills say the antichiral part of each hypermultiplet. Therefore, it acts like
\bea
 \Theta\,: && (H_{11},H_{21},H_{13},H_{23},H_{12},H_{22},H_{14},H_{24})\to \nonumber\\
   &&(\pm H_{11},\pm H_{21},\pm H_{13},\pm H_{23},
      \pm H_{12},\pm H_{22},\pm H_{14},\pm H_{24})
\eea
where the signs $\pm$ indicates the action
\be
{\rm hyper=(chiral,antichiral)} \to  {\rm (chiral},-{\rm antichiral)}
\ee
on the two chiral components of the hypermultiplet. Combining both actions gives
\bea
 \Theta\Theta' : &&  (H_{11},H_{21},H_{13},H_{23},H_{12},H_{22},H_{14},H_{24})
   \to \nonumber \\
   && (\mp H_{11},\mp H_{21},\mp H_{13},\mp H_{23},
      \mp H_{14},\mp H_{24},\mp H_{12},\mp H_{22})
\eea
which is consistent with the fourth term in (\ref{inters}). Therefore, the states which are invariant under both $\Theta$ and $\Theta'$ are precisely two chiral multiplets and there are {\it no further non-chiral} ones. 

A similar analysis can be carried out for the sectors $a_1'a_2$, $a_1,a_3$ and $(a_1'a_3)$, from where also no extra non-chiral matter is found. From the example above it is clear that, if the chiral intersection in the untwisted part is already non-zero (${\cal N}=1$ open string sector) as in the $a_1a_3$ and $a_1'a_3$ sectors, then there cannot be additional non-chiral states. The only thing happening is that some of the chiral fields are projected out by the $\mbb{Z}_2$ actions. Finally, the intersections with the branes of  type $b,c,d$ can be made massive by letting them run through different fixed points.

To summarize, we find an asymptotically free $SU(4)$ gauge theory from the intersecting D-brane model in table \ref{modelnonsplit}. This example proves that, as expected, fractional D6-branes can indeed improve the running of the gauge couplings.

\section{$\cn=1$ and $\cn=0$ chiral flux compactifications}\label{flux}

Up to now, we have been concerned with the construction of type IIA $\cn=1$ chiral models which consist of rigid intersecting D6-branes. As discussed in the introduction, the motivation for these constructions is  finding semi-realistic models without unwanted adjoint fields that spoil asymptotic freedom and, eventually, gauge coupling unification. As we pointed out, this concern can be embedded as part of the well-known problem of moduli stabilization in string theory and, in particular, as the need of stabilizing open string moduli. It thus proves important to combine the present constructions with additional sources of moduli stabilization, such as non-trivial fluxes or analogous backgrounds, in order to conceive a semi-realistic scenario where all the moduli could be lifted. 

The purpose of the present section is to embed our rigid D-brane models in the framework of flux compactifications. Indeed, as noticed in \cite{btl03,cu03}, the model building techniques of intersecting D6-branes can also be implemented in constructing chiral models with RR and NSNS 3-form fluxes. In order to do this, we first need to re-express our type IIA intersecting D-brane models in terms of the equivalent picture of type IIB magnetized D-branes \cite{CB95,bgkl00a,aads00,RR01,cim04}. A prototypical example of this idea has recently been given in  \cite{ms04,ms04a}, where the framework developed in \cite{btl03,cu03} was successfully used in order to build $\cn=1$ chiral flux vacua, as well as $\cn=0$ chiral vacua without RR and NSNS tadpoles.\footnote{See \cite{cl04,cll05} for subsequent models of this kind.}  Notice, however, that these chiral vacua are related to the $\mbb{Z}_2 \ti \mbb{Z}_2$ intersecting D-brane models of \cite{csu01,csu01a}, and hence the D-branes in them are non-rigid or, at best, their moduli are frozen by the background fluxes. As explained in the introduction, it seems more appealing to get rid of such moduli from the very beginning, as rigid D-brane models do.

Hence our strategy will be to consider a type IIB $\T^6/(\mbb{Z}_2 \ti \mbb{Z}_2)$ orientifold background, T-dual to the type IIA $\mbb{Z}_2 \ti \mbb{Z}_2'$ orientifold considered in the previous sections, and introduce RR and NSNS fluxes on top of it. Notice that this possibility has previously been considered in \cite{btl03,ms04a} but, since only bulk D-branes were introduced in those constructions, no $\cn=1$ chiral models could be achieved in this particular background. We will presently overcome such obstruction by introducing fractional D-branes.

\subsection{The model}
\label{typeiibnonsplit}

Following the general strategy described above, as a first step of constructing our chiral flux vacua we will T-dualize the model of table \ref{modelnonsplit} to a type IIB string compactification containing O3-planes. Notice that, by mirror symmetry, we are still dividing  our theory by a $\mbb{Z}_2 \ti \mbb{Z}_2$ orbifold group, but now a $\mbb{Z}_2 \ti \mbb{Z}_2$ without discrete torsion. As a consequence, instead of $3 \ti 16$ collapsed 3-cycles, our orbifold background possesses 48 twisted 2-cycles, which type IIB D$(2p+1)$-branes can wrap.

Our orientifold plane content will be related by T-duality to the one described in sections \ref{intersecting} and \ref{examples}, as well as in \cite{aaads99}, and will contain O3 and O7-planes. More precisely, we will choose to have 64 O3$^{(+,+)}$ and 4 O7$_i^{(-,-)}$ $i=1,2,3$ in our compactification,\footnote{Notice that, since our O3-planes carry positive tension, our construction is slightly different from the general scenario in \cite{gkp01}.} placed on top of the fixed points and tori of the $(\mbb{Z}_2 \ti \mbb{Z}_2) \Om \car$ action. For this to be the case, the geometrical involution $\car$ must be now of the form $\car\, :\, z_I \raw -z_I$.

In order to satisfy RR tadpole cancellation, we now need to introduce type IIB D$(2p+1)$-branes filling up the four non-compact dimensions of the theory. If we restrict to only BPS D-branes, those are given by D3, D7 and $D9-{\overline{D9}}$ pairs. The latter ones are in principle non-BPS, but can be made BPS by introducing suitable magnetic fluxes in their world-volume \cite{btl03,cu03}. In addition, as pointed out in \cite{ms04,ms04a}, such  $D9-\overline{D9}$ pairs can carry ${\overline{D3}}$-brane charge, thus cancelling the O3$^{(+,+)}$-planes RR charges in a supersymmetric fashion.

\begin{table}[htb]
\renewcommand{\arraystretch}{1.5}
\begin{center}
\begin{tabular}{|c||c|c|c||c|}
\hline
 $N_\a$  &  $(n_\a^{1},m_\a^{1})$  &  $(n_\a^{2},m_\a^{2})$
&  $(n_\a^{3},m_\a^{3})$ & type of D-brane \\
\hline
\hline $N_{a_1} = 4$ & $(1,0)$ & $(0,1)$ & $(0,-1)$  & frac. D7$_1$ w/o fluxes \\
\hline $N_{a_2} = 2$ & $(1,0)$ & $(2,1)$ & $(4,-1)$ & frac. D7$_1$ w/ fluxes \\
\hline $N_{a_3} = 2$ & $(-3,2)$ & $(-2,1)$ & $(-4,1)$ & frac. $D9-\overline{D9}$ w/ fluxes \\
\hline
\hline $N_{b}= 2 \ti 2$ & $(1,0)$ &  $ (0,1)$  & $(0,-1)$ & bulk D7$_1$ w/o fluxes \\
\hline
\hline $N_{c}= 2 \ti 4$ & $(0,1)$ &  $ (1,0)$  & $(0,-1)$ &  bulk D7$_2$ w/o fluxes\\
\hline
\hline $N_{d}= 4 \ti 2 N_f$ & $(1,0)$ & $(1,0)$ & $(1,0)$ &  bulk D3 \\
\hline
\end{tabular}
\caption{\small Magnetic numbers of a rigid, $\cn=1$, type IIB chiral vacuum, T-dual to the D-brane model of table \ref{modelnonsplit}. For simplicity we display the sets $b$, $c$ and $d$ as bulk D-branes. In the D-brane description, `w/ fluxes' stands for D$(2p+1)$-branes with internal magnetic fluxes $F_{ab}$ turned on, etc.. \label{Tdualmodelnonsplit}}
\end{center}
\end{table}

We present in table \ref{Tdualmodelnonsplit} an example of a chiral Type IIB flux vacua with rigid D-branes of the above kind. Actually, this model is nothing but a T-dual version of the one constructed in Section \ref{examples}, table \ref{modelnonsplit}. Instead of rigid D6-branes wrapping 3-cycles, we now deal with D3, D7 and D9-branes with magnetic fluxes. Instead of the previous wrapping numbers, the topological data of the configuration are now given by magnetic quantum numbers, which specify which kind of D$(2p+1)$-branes are we dealing with and which are the gauge instanton numbers in its woldvolume. We will not give here a definition of these magnetic numbers and of the conventions associated with them. These can, nevertheless, be easily inferred from the usual T-duality rules and the D-brane description of table \ref{Tdualmodelnonsplit}. For a more detailed description of these kind of models we refer the reader to \cite{btl03,cu03}.

Being a construction T-dual to the one in table \ref{modelnonsplit}, this D-brane model shares the same low energy spectrum  and physical properties as the one previously constructed. In particular, twisted RR tadpoles are cancelled separately by each set $a,b,c,d$ of fractional D-branes. In addition, all the untwisted RR tadpoles, with the exception of the D3-brane charge, are canceled by the above D-brane content. The D3-brane tadpole can be satisfied by suitably choosing $N_f$ in table \ref{Tdualmodelnonsplit}. However, when we introduce background fluxes in the next subsection they will also contribute to the D3-brane tadpole, so we will wait until then before fixing $N_f$.

Just as before, the model can be made $\cn=1$ supersymmetric by tuning the moduli of the compactification. In the present type IIB context, this implies a tuning of the K\"ahler moduli of the compactification. More precisely, the conditions (\ref{susynonsplit}) translate into 
\be
\begin{array}{c}\vspace*{.2cm}
2 \ca_2 = \ca_3, \\
\arctan \left(2\ca_1/3\right) + \arctan \left(\ca_2/2\right) + \arctan \left(\ca_3/4\right) = \pi,
\end{array}
\label{Tdualsusynonsplit}
\ee
where $\ca_i$ is the area of the $i^{th}$ $\T^2$. In addition, one needs to keep each of the twisted sectors of the compactification as collapsed 2-cycles, at least those combinations under which the fractional D-branes of the model are non-trivially charged. Any deviation from these particular points in the K\"ahler moduli space will be seen, in the D-brane effective theory, as a non-vanishing FI-term \cite{km99,cim02}. Hence, we expect SUSY to be restored after some particular set of charged open string moduli have received  a VEV.

\subsection{Adding fluxes}

Let us now briefly discuss how we can introduce non-trivial background fluxes in the present context. Following \cite{btl03,cu03}, we introduce both RR and NSNS 3-form field strength backgrounds $F_3$ and $H_3$ with non-trivial components on the internal dimensions of the compactification. Consistency of the compactification imposes that these 3-forms satisfy the Bianchi identities $dF_3 = dH_3 = 0$ which, in the present context, can be easily satisfied by requiring $F_3$, $H_3$ to be constant 3-forms in the untwisted cohomology of $\T^6/(\mbb{Z}_2 \ti \mbb{Z}_2)$. In addition, these 3-form fluxes must be properly quantized over any 3-cycle of the orientifolded geometry $\T^6/\G$. In the case of toroidal orientifolds, the quantization conditions amount to
\be
{1 \over (2 \pi)^2 \a^\prime} \int_{\Sigma} F_3 \in N_{\rm min} \mbb{Z}, \quad \quad
{1 \over (2 \pi)^2 \a^\prime} \int_{\Sigma} H_3 \in N_{\rm min} \mbb{Z}
\label{quant}
\ee
where $\Sigma$ is a 3-cycle on the covering space ${\bf T}^6$, and $N_{\rm min}$ is an integer number  depending on the orbifold group $\G$. For $\G = \mbb{Z}_2 \ti \mbb{Z}_2$ without discrete torsion, we find $N_{\rm min} = 4$ \cite{btl03}. 

This non-trivial background is usually encoded in the language of type IIB supergravity, in terms of the complexified 3-form flux
\be
G_3 = F_3 - \tau H_3
\label{G3}
\ee
where $\tau = a + i/g_s$ is the usual axion-dilaton coupling. By looking at the supergravity action it is easy to see that these fluxes both gravitate and couple to the $C_4$ RR potential, hence carrying both D3-brane charge and tension. In D3-brane units, the RR charge is given by the topological quantity
\be
N_{\rm flux} = {1 \over (4\pi^2 \a^\prime)^2} \int_{\cm_6} H_3 \wedge F_3 = {i \over (4\pi^2 \a^\prime)^2} \int_{\cm_6} {G_3 \wedge \overline{G}_3 \over 2\pim \tau}
\label{RRcharge}
\ee
where $\cm_6$ is the six-dimensional manifold on which we compactify our theory. If we take $\cm_6 =\T^6$ and impose the quantization conditions (\ref{quant}), then we obtain that $N_{\rm flux}$ is an integer number proportional to $(N_{\rm min})^2$. In addition, absence of NSNS tadpoles implies that $N_{\rm min} \geq 0$. These facts are quite relevant for model building purposes. Indeed, the D3-brane RR tadpole condition gets modified when we consider non-trivial fluxes, and hence the first constraint in (\ref{RRtad1}) becomes
\be
\sum_a N_a n_a^1 n_a^2 n_a^3 + \half N_{\rm flux}  =  -16,
\label{RRD3flux}
\ee
with the rest of the tadpole conditions remaining the same.

In addition, a $G_3$ flux will carry some non-trivial D3-brane tension $T_{\rm flux}$. Such tension is not a topological quantity, but rather depends on the complex structure moduli and complex dilaton. From the $D=4$ effective theory viewpoint, the excess of tension $T_{\rm flux}$ can be seen as a scalar potential $V_{sc}$ for the above moduli, which can be derived from a Gukov-Vafa-Witten superpotential \cite{gvw99}. In the present context, the minimum of $V_{sc}$ is reached whenever $G_3$ attains the ISD condition, that is, when it satisfies $*_6 G_3 = i G_3$. Again, since the ISD condition is only satisfied for a particular set of values of the complex structure moduli and complex dilaton, most of these moduli get lifted by introducing the 3-form flux \cite{drs99,gkp01}.

In practice, whenever we construct a D-brane configuration  preserving $\cn=1$ supersymmetry by adding an ISD $G_3$ flux with vanishing RR tadpoles, the string vacuum  is free of NSNS tadpoles and has vanishing $D=4$ cosmological constant.\footnote{At least at the supergravity level.} This does not mean, however, that our compactification is $\cn=1$ supersymmetric. Indeed, in order to preserve $\cn=1$ supersymmetry $G_3$ not only needs to be  ISD, but also  must consist of only primitive (2,1)-forms \cite{gp00}. If, on the contrary, $G_3$ contains a non-trivial (0,3) component, then supersymmetry will be broken by means of a non-vanishing F-term for the overall K\"ahler modulus. In this case, the $D=4$ gravitino will acquire a mass proportional to the flux density \cite{gkp01}, and soft terms will be induced on the D-brane sector of the theory \cite{lrs04,ciu04}.

Given this general scenario, let us now consider the particular D-brane model of table \ref{Tdualmodelnonsplit}. Given the quantization conditions and that $G_3$ is ISD, we can write $N_{\rm flux} = n \cdot 16$, $n \in \mbb{N}$. The D3-brane RR tadpole condition (\ref{RRD3flux}) then reads
\be
N_f + n = 2,
\label{RRD3fin}
\ee
which has several solutions. For instance, choosing $N_f=2$, $n=0$ corresponds to a model without 3-form fluxes, which is T-dual to the type IIA D6-brane model of table \ref{modelnonsplit}. Alternatively, $N_f=n=1$ corresponds to a vacuum where some D3-branes are present and a non-trivial $G_3$ flux is turned on. Finally, $N_f=0$, $n=2$ corresponds to a solution where all D3-branes have been replaced by the $G_3$ flux. 

Let us focus on this last possibility. Notice that one possible flux satisfying the ISD condition is given by
\be
G_3\, =\,  2i \, ( d{\overline z}_1dz_2dz_3 + dz_1d{\overline z}_2dz_3).
\label{susyflux}
\ee
Such 3-form flux constrains the complex structure moduli $\tau_i$, $i=1,2,3$, of the $i^{th}$ $\T^2$ factor and the complex dilaton $\tau$ to satisfy $\tau_1 \tau_2 = \tau_3 \tau = -1$, fixing half of these complex moduli. Moreover, its non-vanishing components are exclusively (2,1)-forms, so that this flux preserves the $\cn=1$ supersymmetry of the orientifold background. Hence, we have succeeded in constructing an $\cn=1$ chiral string theory vacuum with two different sources of moduli stabilization. On the one hand, most positions/Wilson lines of the D-brane sector are projected out by the orientifold action, receiving a string scale mass. On the other hand, half of the complex structure/dilaton moduli get lifted by means of the flux-induced scalar potential, receiving a mass of the order 
\be
M_{flux}\simeq {\a'\over \sqrt{{\rm Vol}(\cm_6)}}\simeq {M_s^2\over M_{pl}}. 
\ee
Finally, we also expect the $G_3$ flux to induce  a mass of the same order to the fields which correspond to the position moduli of the D7-branes $b$ and $c$ \cite{lrs04,ciu04}.

Another possibility is introducing the flux
\be
G_3\, =\,  d{\overline z}_1dz_2dz_3 + dz_1d{\overline z}_2dz_3 + dz_1dz_2d{\overline z}_3 + d{\overline z}_1 d{\overline z}_2 d{\overline z}_3.
\label{nonsusyflux}
\ee
which, having a non-vanishing (0,3) component, breaks supersymmetry down to $\cn=0$. Following the computations of \cite{ms04a}, it is easy to see that the scalar potential generated by (\ref{nonsusyflux}) vanishes if and only if the complex moduli $\tau_i$ and $\tau$ are constrained to be pure imaginary and to satisfy $\pim \tau_1 \pim \tau_2 \pim \tau_3 = \pim \tau$. This background flux not only stabilizes complex structure and D7-brane position moduli, but it also gives a mass to the gravitino and, moreover, to the scalar components of the $\cn=1$ chiral multiplets in table \ref{chiralnonsplit}. Thus, in principle, this framework allows to lift open string moduli associated to both non-chiral and chiral matter in the low energy theory. Of course, the absence of the latter is linked to the pattern of SUSY-breaking soft-terms which appear in the low energy effective theory, and it would be interesting to study the phenomenological consequences of such soft terms in the present model.\footnote{This kind of analysis have been performed in \cite{LI04,lrs04a,kklw04,if04}, for the particular case of the MSSM-like local model constructed in \cite{cim03}.} 

In principle, one could also think that this $\cn=0$ flux lifts most K\"ahler moduli in the compactification in an indirect way. Indeed, similarly to \cite{cgqu03,if04} one could argue that the open string chiral fields participate in a D-term potential where the twisted and untwisted K\"ahler moduli are also involved. Since the scalar components of such chiral multiplets acquire a mass by effect of the $\cn=0$ flux, then the degeneracy of the D-term potential gets lifted, and the FI-terms of the theory are dynamically fixed to vanish. Now, since the FI-terms depend on the K\"ahler moduli of the compactification, and in particular on the deviation of conditions such as (\ref{Tdualsusynonsplit}), the absence of FI-terms would translate into new constraints for the K\"ahler moduli, which are hence also lifted. One must bear in mind, however, that this kind of reasoning relies on partial effective field theory results, in which F and D-term potentials arise from apparent different sources. A more complete analysis would require the lift of this type IIB compactifications to F-theory, along the lines of \cite{lmrs05}, where both potentials arise from the same 4-form background flux.

Finally, let us point out that, whenever $N_f \neq 2$ the spectrum of table \ref{chiralnonsplit} presents chiral anomalies. In particular, mixed anomalies where the $U(1)$ of $U(2)_{a_3}$ is involved. As explained in \cite{AU02b}, such anomalies are cured by the a 3-form flux satisfying (\ref{RRD3fin}), for  a Wess-Zumino term contributing to that anomaly is generated by $G_3$. Such kind of flux-induced Wess-Zumino terms were considered in \cite{aiu02} in order to solve the strong CP problem. It would be interesting to address further phenomenological consequences of this fact in the present context.

\section{Conclusions}\label{conclusions}

In this paper we have discussed the phenomenologically important issue of freezing open-string moduli in the framework of intersecting D-brane models, with particular emphasis on ${\cal N}=1$ chiral constructions.  Although by now there exist many intersecting D-brane models whose chiral spectrum is remarkably close to the Standard Model, there are generically additional non-chiral states in their low energy spectrum which pose phenomenological challenges for these models. 

Such unwanted non-chiral states can be seen as geometric open string moduli associated with the cycles that the D-branes wrap around. For instance, if we consider  a stack of D6-branes whose world-volume contains a $U(N)$ gauge symmetry, these moduli are massless chiral superfields in the adjoint representation, and correspond geometrically to the positions and  Wilson lines of such D-branes. To get rid of these open string moduli, one must develop a framework to construct intersecting D-brane models where the associated D-branes are wrapped around rigid cycles. As illustrative  examples of such D-brane models with rigid 3-cycles, we have constructed some supersymmetric intersecting D6-brane vacua on the $\mbb{Z}_2\times \mbb{Z}_2$ orbifold with discrete torsion. We presented several explicit chiral supersymmetric D-brane models with frozen open string moduli. The simplest class of such models involves bulk branes splitted into fractional branes, and generically lead to multiple gauge factors such as  $U(N)^2$ or $U(N)^4$. A more interesting class of chiral vacua contains fractional D-branes that are not arranged on the regular representation of $\mbb{Z}_2 \ti \mbb{Z}_2$. However, models with such D-branes are more difficult to construct because of a large number of RR tadpole constraints that need to be satisfied. We presented a strategy to systematically find such models. We also provided two $\cn=1$ chiral examples, one of which contains the realistic features of the Pati-Salam model. Moreover, we have discussed some phenomenological features of these construction, such as asymptotic freedom and gaugino condensation. Finally, we described how the T-dual Type IIB version of the models constructed here can be embedded in the framework of flux compactification, where the 3-form background flux provides a source to stabilize closed string moduli complementary to the effects of rigid cycles in freezing open string moduli.

Although the models presented are not fully realistic, they provide some simple examples to illustrate how the general formalism developed here can be applied. For instance, we have limited our search to models with untilted tori and, as a result, only 4 generation models have been found. This should not be viewed as an obstruction to obtain realistic models in this context. On the contrary, just as in the $\mbb{Z}_2 \times \mbb{Z}_2$ construction of \cite{csu01,csu01a}, once tilted tori are allowed, more realistic spectra can presumably be obtained. In this same spirit, it would also be interesting to perform a systematic search for realistic models based on rigid fractional branes, where the SM sector of such models can be embedded into Pati-Salam type models (as in the examples presented here) or GUT models with $SU(5)$ gauge symmetries. Complementary to searching for more realistic vacua, one could also perform a statistical analysis of the set of solutions to the tadpole cancellation conditions, in the same spirit as in \cite{bghlw04}. We hope to report on the results of these studies in the future.

 \vskip 1cm
 {\noindent  {\Large \bf Acknowledgments}}
\vskip 0.5cm 

We would like to thank  Angel Uranga for collaboration at an early stage of this work and for useful discussions. 
RB, MC, and GS are grateful to the  DAMTP at  Cambridge for hospitality. RB also thanks PPARC for financial support during part of this project. 
The work of MC was supported in part by  the DOE grant DOE-EY-76-02-3071, NSF grant INT02-03585, and Fay R. and Eugene L. Langberg endowed Chair.
FM and GS were supported in part by NSF CAREER Award No. PHY-0348093,
DOE grant DE-FG-02-95ER40896, and a Research Innovation Award from Research Corporation. FM and GS also thank the Perimeter Institute for Theoretical Physics
for hospitality while part of this work was done.

\newpage

\section{Appendix: K-theory constraints}

As pointed out in \cite{AU00}, cancellation of RR tadpoles in orientifold compactifications usually imposes stronger constraints than the absence of divergences in one-loop open string diagrams. From imposing the absence of these RR divergences, one usually recovers a set consistency constraints which can be encoded in the homology of the compact manifold, and which reduce, in the supergravity limit, to the consistency of the Bianchi identities/equations of motion of the $D=10$ RR fields. In general, these supergravity constraints can be recasted as the cancellation of a total homological RR charge that D-branes and O-planes carry, just as in (\ref{RRCY}). It turns out, however, that in orientifold compactifications D-branes may also carry torsion charges which are invisible to homology. These torsion charges are usually $\mbb{Z}_2$-valued and, although more difficult to characterize than the familiar homological charges, we also need to impose their cancellation in order to build a consistent string compactification.

In this appendix we derive these extra consistency constraints imposed by K-theory on the $\mbb{Z}_2 \ti \mbb{Z}_2'$ orientifold considered in the main text. More precisely, we restrict to the case where the crosscap charges of the orientifold are chosen to be $\eta_{\Om\car} = -1$ and $\eta_{\Om\car\Th} = \eta_{\Om\car\Th'} = \eta_{\Om\car\Th\Th'} = 1$, which is the case considered in Section \ref{examples}. By the same token, we restrict to a complex structures given by square $\T^2$ factors, i.e., such that $\b^I = 0, \forall I$. 

As explained in \cite{AU00}, uncanceled non-homological K-theory charges usually do not show up as chiral anomalies in the low energy field theory, just as more familiar RR uncanceled charges would do. On the other hand, one can see their effect by introducing suitable D-brane probes in the theory, and looking at the chiral anomalies that may develop on the world-volume of such probes. A D-brane model free of uncanceled K-theory charges would not develop a chiral anomaly on any possible D-brane probe that we may introduce. In particular, one can think of introducing D-brane probes such that they yield a $USp(2) \simeq SU(2)$ gauge group in their world-volume theory. Although free of cubic anomalies, an $SU(2)$ gauge group suffers from a global gauge anomaly if there is an odd number of $D=4$ chiral fermions transforming in the fundamental representation \cite{EW82}.

We then need  to require the absence of $SU(2)$ global anomalies for any D6-brane probe that we may introduce in the $\mbb{Z}_2 \ti \mbb{Z}_2'$ background specified above. As discussed in the main text, there are exactly 4 $\ti$ 64 fractional D-branes which may develop such a gauge group, namely those with bulk wrapping numbers $(1,0)(1,0)(1,0)$ and arbitrary $\eps^g$'s. Thus, we would expect 256 different ways of generating $SU(2)$ global anomalies and hence up to 256 extra consistency conditions. In fact, requiring the absence of such global anomalies translates into conditions of the form
\be
\begin{array}{lcr}\vspace*{.2cm}
\sum_a {1 \over 4} N_a \left[ m_a^1 m_a^2 m_a^3 + Q_1^a m_a^1 + Q_2^a m_a^2 + Q_3^a m_a^3 \right] & \in & 2 \mbb{Z}\\ \vspace*{.2cm}
\sum_a {1 \over 4} N_a \left[ m_a^1 m_a^2 m_a^3 + Q_1^a m_a^1 - Q_2^a m_a^2 - Q_3^a m_a^3 \right]& \in & 2 \mbb{Z} \\ \vspace*{.2cm}
\sum_a {1 \over 4} N_a \left[ m_a^1 m_a^2 m_a^3 - Q_1^a m_a^1 + Q_2^a m_a^2 - Q_3^a m_a^3 \right] & \in & 2 \mbb{Z}\\ \vspace*{.2cm}
\sum_a {1 \over 4} N_a \left[ m_a^1 m_a^2 m_a^3 - Q_1^a m_a^1 - Q_2^a m_a^2 + Q_3^a m_a^3 \right] & \in & 2 \mbb{Z}
\end{array}
\label{Ktheory}
\ee
where the $m_a$'s stand for the bulk wrapping numbers of the D-brane stack $a$. On the other hand, $Q^a_i$ are integer numbers that depend on the specific 3-cycle $\Pi_a$ that the D-brane is wrapping. More precisely,
\be
Q_i^a\, =\, \left\{
\begin{array}{ccl}\vspace*{.1cm}
1 & {\rm if} & m_a^j m_a^k \equiv 1\, {\rm mod}\, 2 \\\vspace*{.1cm}
2\ {\rm and}\ 0 & {\rm if} & m_a^j m_a^k \equiv 0 \, {\rm mod}\, 2 \quad {\rm and} \quad  m_a^j+m_a^k \equiv 1\,  {\rm mod}\, 2 \\
4\ {\rm and}\ 0 & {\rm if} & m_a^j m_a^k \equiv 0 \, {\rm mod}\, 2 \quad {\rm and} \quad  m_a^j+m_a^k =0 \,  {\rm mod}\, 2 
\end{array}
\right.% \quad  i \neq j \neq k \neq i 
\label{Qs}
\ee

Where, in the case that $m_a^j m_a^k \in 2\mbb{Z}$, the choice of $Q^a_i$'s depends on the particular K-theory charge that we are considering and, in general, we need to consider all possible combinations to scan all K-theory charges.

From this, we see that the `K-theoretical' RR tadpole cancellation can be quite constraining in these models. In particular, any D-brane which gives a non-trivial (odd) contribution to the sums (\ref{Ktheory}), for some choice of the $Q$'s, will posses a non-homological K-theory charge. One can check that any fractional D6-brane will have non-trivial K-theory charges unless their wrapping numbers are of the form
\be
(\rm{odd}, \rm{even}) \otimes (\rm{odd}, \rm{even}) \otimes (\rm{odd}, \rm{even}),
\label{noKap}
\ee
On the other hand, in order to avoid non-vanishing $\mbb{Z}_2$ charges we can consider stacks with an even number of D6-branes. That is, $N_a \in 2\mbb{N}$.

Finally, if we consider  bulk D6-branes in our model then the integers $Q_i^a$ above must be taken to be 0. A model with only bulk D-branes will have as extra K-theory constraints
\be
\sum_a N_a m_a^1 m_a^2 m_a^3 \, \in \, 8\mbb{Z}
\label{Ktheorybulk}
\ee
where, as before, $N_a$ is counting fractional D-branes, so it will always come in multiples of four in the case of bulk D-branes. From (\ref{Ktheorybulk}) we then recover the conditions deduced in \cite{ms04a}.

\clearpage
\nocite{*}
\bibliography{rev2}
\bibliographystyle{utphys}

\end{document}